\documentclass[fleqn,usenatbib]{mnras}
\usepackage{newtxtext,newtxmath}
\usepackage[T1]{fontenc}
\usepackage[normalem]{ulem}
\usepackage{epsfig}
\usepackage{mathtools}
\usepackage{cuted}

\title[Collisional heating of icy planetesimals]{Collisional heating of icy planetesimals. I. Catastrophic collisions}

\author[Bj\"{o}rn J. R. Davidsson]{
Bj\"{o}rn J. R. Davidsson,$^{1}$\thanks{E-mail: bjorn.davidsson@jpl.nasa.gov}
\\
$^{1}$Jet Propulsion Laboratory, California Institute of Technology,  M/S 183--601, 4800 Oak Grove Drive, Pasadena, CA 91109, USA
}

\date{Accepted 2023 February 28. Received 2023 February 23; in original form 2022 October 30}

\pubyear{2023}

\begin{document}
\label{firstpage}
\pagerange{\pageref{firstpage}--\pageref{lastpage}}
\maketitle


\begin{abstract}
Planetesimals in the primordial disc may have experienced a collisional cascade. If so, the comet nuclei later placed in the Kuiper belt, scattered disc, and Oort Cloud would primarily be fragments and collisional rubble piles from that cascade. However, the heating associated with the collisions cannot have been strong enough to remove the hypervolatiles that are trapped within more durable ices, because comet nuclei are rich in hypervolatiles. This places constraints on the diameter of the largest bodies allowed to participate in collisional cascades, and limits the primordial disc lifetime or population size. In this paper, the thermophysical code \textsc{nimbus} is used to study the thermal evolution of planetesimals before, during, and after catastrophic collisions. The loss of CO during segregation of $\mathrm{CO_2:CO}$ mixtures and during crystallisation of amorphous $\mathrm{H_2O}$ is calculated, as well as mobilisation and internal relocation of $\mathrm{CO_2}$. If an amorphous $\mathrm{H_2O}$ host existed, and was protected by a $\mathrm{CO_2:CO}$ heat sink, only diameter $D<20\,\mathrm{km}$ (inner disc) and $D<64\,\mathrm{km}$ (outer disc) bodies could have been involved in a collisional cascade. If $\mathrm{CO_2}$ was the only CO host, the critical diameters drop to $D<20$--$32\,\mathrm{km}$. Avoiding disruption of larger bodies requires a primordial disc lifetime of <9 Myr at 15 au and <50--70 Myr at 30 au. Alternatively, if a 450 Myr disc lifetime is required to associate the primordial disc disruption with the Late Heavy Bombardment, the disc population size must have been 6--60 times below current estimates.
\end{abstract}


\section{Introduction} \label{sec_intro}

Current--day comets, Centaurs, trans--Neptunian objects (TNOs), Oort Cloud objects, and/or their planetesimal ancestors, were once 
located in the primordial disc. This ancient structure is thought to have stretched from $\sim 15\,\mathrm{au}$ to $\sim 30\,\mathrm{au}$ from the Sun, 
being located exterior to an initially compact giant--planet orbital configuration \citep{gomes04}. The primordial disc was disrupted by a gravitational 
instability among the giant planets. This led to the formation of the present--day reservoirs of small icy bodies: the dynamically hot Kuiper belt 
(superimposed on top of, and beyond,  a small pre--existing population of dynamically cold objects at 42--$47\,\mathrm{au}$), the scattered disc, and the 
Oort Cloud \citep{gomesetal05b,morbidellietal05,tsiganisetal05,levisonetal08}. Objects in the last two populations are prone to dynamical evolution that allow 
some to enter the inner Solar system as Centaurs (some of which evolve dynamically to become Jupiter Family comets, or JFCs), Halley Type comets, or 
dynamically new comets \citep{fernandez80,duncanlevison97,brassermorbidelli13}.

The thermal processing of primordial disc objects prior to their dispersal was investigated by \citet{davidsson21}. He found that objects in the $D=4$--$200\,\mathrm{km}$ 
diameter range lose all their pure CO ice on time--scales ranging $70\,\mathrm{kyr}$--$13\,\mathrm{Myr}$ 
(depending on size) due to protosolar and radiogenic heating by long--lived isotopes. These time--scales are shorter than most estimates of the primordial disc 
lifetime (15--$450\,\mathrm{Myr}$; see below). Therefore, pure CO (and other hypervolatiles such as $\mathrm{N_2}$, $\mathrm{O_2}$, $\mathrm{CH_4}$, 
$\mathrm{C_2H_6}$, and noble gases) are not expected in small objects at the time of their dispersal towards colder regions. \citet{choietal02} also found 
complete CO loss from a $D=200\,\mathrm{km}$ model object at heliocentric distance $r_{\rm h}=30\,\mathrm{au}$ on a $\sim 10\,\mathrm{Myr}$ timescale, though they 
only modelled the heat conduction process and not the gas diffusion process. Significant CO loss  at $r_{\rm h}=41$--$45\,\mathrm{au}$ was reported by 
\citet{desanctisetal01} in concurrent heat and gas diffusion modelling similar to that of \citet{davidsson21}, though they did not follow the process until completion. \citet{steckloffetal21} and \citet{lisseetal21,lisseetal22} 
used a surface--ice sublimation model, and the time--scale of the thermal wave to reach the core, to argue that Arrokoth and other small to medium--sized Kuiper belt objects 
ought to be hypervolatile--free. \citet{prialnik21} report complete loss of CO and $\mathrm{CH_4}$ from an Arrokoth analogue body at $r_{\rm h}=44\,\mathrm{au}$ in 
$100\,\mathrm{Myr}$ (but survival at $r_{\rm h}\stackrel{>}{_{\sim}}200\,\mathrm{au}$), though computational details in her conference abstract are sparse. Furthermore, 
the activation distance of most comets excludes hypervolatiles stored as clean ice \citep{jewitt09}.

However,  CO outgassing from comets is ubiquitous in the inner Solar system \citep[abundance ratios range $0.002\leq \mathrm{CO/H_2O}\leq 0.294$ 
and have a mean of $\langle\mathrm{CO/H_2O}\rangle=0.063$ for a sample of 21 comets;][]{ahearnetal12}. Because pure CO ice cannot have survived the primordial 
disc stage in most small objects, a fraction of the CO must have been stored within a less volatile medium, such as amorphous 
$\mathrm{H_2O}$ \citep[e.~g.,][]{prialnikandbarnun87,prialnikandbarnun88,prialniketal04,prialniketal08,jewitt09} or $\mathrm{CO_2}$ \citep{gascetal17,davidsson21}. 

These hypervolatile reservoirs are fragile and sensitive to further disturbances. The ambient planetesimal core temperature in the outer half of the primordial 
disc would typically have been near $\sim 50\,\mathrm{K}$. In order to initiate  $\mathrm{CO_2:CO}$ segregation ($\sim 63\,\mathrm{K}$) and amorphous 
$\mathrm{H_2O}$ crystallisation ($\sim 85\,\mathrm{K}$), global long--term temperature elevations of merely $\sim 10$--$35\,\mathrm{K}$ are sufficient. One 
potential source of such heating is catastrophic collisions among icy planetesimals, a process that \citet{davidsson21} did not consider. It is clear that a potential collisional 
cascade cannot have been sufficiently energetic to set off wide--spread segregation and crystallisation. If that happened, the last reservoirs of CO would be lost, and the 
hypervolatile--free/poor fragments (and re--accumulated rubble piles thereof) would not be good comet analogues. 

The current paper is therefore devoted to the problem of investigating the degree of CO loss during catastrophic collisions (taken as a representative of the 
loss of all hypervolatiles). Bodies that are large enough that their energetic catastrophic disruptions lead to massive CO loss, could not have participated in 
a collisional cascade. Thermophysical simulations thereby become powerful tools for identifying the largest body (with diameter $D_{\star}$) that could have been commonly involved in 
collisional cascades, because of the necessity of preserving significant deposits of CO. Such investigations have the potential of placing novel constraints on the 
lifetime and number density of the primordial disc (i.~e., the disc must have disrupted sufficiently early, and/or the number density must have been sufficiently 
low, to prevent collisions that lead to massive hypervolatile loss). In the following, I argue that such constraints are urgently needed.

\emph{The primordial disc lifetime.} A lifetime consistent with the Late Heavy Bombardment at 
$\sim 450\,\mathrm{Myr}$ \citep{morbidellietal12,marchietal13b} is dynamically possible if the giant planets emerged on resonant orbits when the stabilising 
gas disc evaporated, and the gap between the outermost giant planet and the inner disc edge was sufficiently large \citep{gomesetal05b,morbidellietal07,levisonetal11}. 
However, the giant--planet migration associated with such instabilities risks to cause an unacceptable level of terrestrial--planet orbital excitation 
through secular resonant coupling \citep{agnorandlin12}. One way to overcome such difficulties is the `jumping Jupiter' scenario, where a late 
instability remains possible if the orbital separation of Jupiter and Saturn increases abruptly as they gravitationally scatter one or several ice giants 
\citep{morbidellietal09c,brasseretal09}. A second possibility is that the gravitational instability is unrelated to the Late Heavy Bombardment, 
and that it took place prior to terrestrial planet formation, so that the excited embryos re--establish orbits with low eccentricity and 
inclination through dynamical friction against remaining planetesimals before growing further \citep{agnorandlin12}. The average time needed to grow 
terrestrial embryos to half their final size is about 15--$25\,\mathrm{Myr}$ according to \citet{obrienetal06}, suggesting that the instability 
may have occurred as early as that. Various attempts to constrain the timing of the instability have been made by considering statistically large samples of models, and searching 
for cases that best reproduce various Solar system constraints. \citet{nesvornyandmorbidelli12} studied cases with up to six giant planets 
(of which 1--2 ice giants are ejected from the Solar system), and found satisfactory solutions for instability times ranging 3.2--$34\,\mathrm{Myr}$. 
\citet{desousaetal20} found mean instability times ranging 37--$62\,\mathrm{Myr}$. Although giant planet migration in the gas disc often place the 
planets on stable resonant orbits \citep[e.~g.,][]{morbidellicrida07,morbidellietal07}, \citet{desousaetal20} studied cases where the giant planets 
emerge on unstable orbits after gas disc evaporation. In such cases, they found mean instability times as short as $4\,\mathrm{Myr}$. I note, that 
\citet{nesvorny18}, in his comprehensive review of the problem, favours entry of Neptune into the primordial disc `a few tens of millions of years 
after the dispersal of the protosolar nebula'.

However, even if the instability occurred very early, the disruption of the primordial disc is not instantaneous. In order to reproduce the inclination distribution 
of Kuiper belt objects, \citet{nesvorny15} found that the passage of Neptune through the disc needs to proceed for at least $10\,\mathrm{Myr}$. All in all, it is 
therefore likely that the primordial disc lifetime ranged 15--$450\,\mathrm{Myr}$.

\emph{The primordial disc population size}. Estimating the number of objects that populated the disc is difficult. This problem stems from the difficulty of determining the fraction of the objects 
ending up in the scattered disc and Oort Cloud, evaluating the losses in those populations during the Solar system lifetime, estimating the fractional rates of injection 
towards the inner Solar system,  and determining the current sizes of comet reservoirs through the influx of dynamically new comet (Oort Cloud), or through the influx 
of JFCs and deep observational surveys (scattered disc), while having to deal with low--number statistics and various forms of observational bias 
\citep[for a brief review, see e. g., section~4.5 in][]{davidssonetal16}. Estimates of the primordial disc content of $D\geq 2\,\mathrm{km}$ bodies ranges 
from $3\cdot 10^9$--$2\cdot 10^{10}$ \citep{bottkeetal12} to $(1.9$--$5)\cdot 10^{11}$ \citep{morbidellietal09,brassermorbidelli13}. Similar two--orders--of--magnitude 
discrepancies of the scattered disc population size exist amongst various predictions and attempts to determine it observationally  \citep{volkmalhotra08}. 

If the substantial ranges on primordial disc mass and lifetime could be narrowed down, by finding thermophysical 
constraints in addition to the dynamical and observational ones, our understanding of the early Solar system evolution would increase substantially. It would have implications on how 
and when the giant planets migrated, the time available for the growth of Pluto-- and Eris--sized bodies, and the timing of water and organics injection into the inner Solar system 
relative the growth sequence of terrestrial planets and the emergence of Life.

Finally, primordial disc mass and lifetime affect the way we view the scientific contributions of cometary studies. If the highest estimates of the primordial disc population size are accurate, 
comets like the \emph{Rosetta} target 67P/Churyumov--Gerasimenko (hereafter, 67P) would necessarily have to be considered collisional fragments or rubble piles \citep{rickmanetal15,morbidellirickman15}, particularly 
if the primordial disc lifetime was long. However, if the population size was small and/or the lifetime was short, there might not have been many destructive collisions and most comets 
could be primordial \citep{massironietal15,davidssonetal16}. Resolving this issue would tell us whether the physical properties of comets, revealed by spacecraft missions, primarily inform 
about early solar nebula accretion processes, or about later secondary processing through destructive collisions. Because the potential collisional processing would have taken 
place before the dislocation of objects to the Oort Cloud, this question is also highly relevant for \emph{Comet Interceptor} and similar missions targeting the presumably 
primitive dynamically new comets.

This paper is structured as follows: section~\ref{sec_model} summarises the thermophysical model used for this work, and section~\ref{sec_prep} describes 
some necessary preparatory work. In particular, section~\ref{sec_prep_largest_parent} identifies the largest potential parent body to be considered in this 
paper, section~\ref{sec_prep_coll_environ} describes the collisional environment, section~\ref{sec_prep_density_wasteheat} discusses the generation of 
waste heat in collisions, and section~\ref{sec_prep_method} summarises the methodology for the thermophysical simulations. The main results are presented in 
section~\ref{sec_results}, they are discussed in section~\ref{sec_discussion}, and the conclusions are summarised in section~\ref{sec_conclusions}. Furthermore, 
Appendix~\ref{appendix01} discusses the problem of heating in continuum mechanics collision codes.

\section{The thermophysical model} \label{sec_model}

The modelling work is here made with the ``Numerical Icy Minor Body evolUtion Simulator'', or \textsc{nimbus}. The code is described in 
full detail by \citet{davidsson21}, therefore only a brief summary is made here, focusing on the applied model parameters. \textsc{nimbus} has also been used to model 
different aspects of Comet 67P \citep{davidssonetal21,davidssonetal22,davidssonetal22b}, and of sporadically active Asteroid (3200)~Phaethon \citep{masieroetal21}.

\textsc{nimbus} considers a spherical body and allows for any temporally changing orbit and spin state to be considered. Here, circular orbits in 
the ecliptic plane with semimajor axes $\{15,\,23,\,30\}\,\mathrm{au}$ are considered, assuming a fixed $\{\lambda,\,\beta\}=\{0^{\circ},\,45^{\circ}\}$ spin 
axis orientation (ecliptic longitude and latitude) in fast--rotator mode. The model bodies are resolved by 18 angular--equidistant latitudinal slabs, and 
radially by 87--147 cells, with widths growing from $5\,\mathrm{m}$ at the surface to 0.4--$1.5\,\mathrm{km}$ at the core, depending on body size 
(diameters $D=\{16,\,20.2,\,25.4,\,32,\,40.3,\,50.8,\,64\}\,\mathrm{km}$ are considered). See section~\ref{sec_prep_largest_parent} for further discussion on 
body sizes. The luminosity time evolution of the protosun follows that of a $1\,\mathrm{M_{\odot}}$ star \citep{pallastahler93}, but a Solar nebula 
clearing time of $t_{\rm c}=3\,\mathrm{Myr}$ is applied to limit the luminosity to $\stackrel{<}{_{\sim}} 1$ times the current one. The thermal processing of the model bodies is 
primarily determined by the collisional energy injection, therefore parameters that are fine--tuning solar heating (shape, orbital eccentricity and inclination, spin--axis orientation, 
fast--rotator assumption) have negligible influences on the $D_{\star}$ estimate. Shape does influence cooling times (by providing a different surface area available for 
radiative cooling compared to the volume of collisionally heated material, than a sphere). However, shape has no influence on the core temperature maximum following collisional 
flash--heating, that determines whether the segregation or crystallisation thresholds are reached (and whether CO is massively lost).

The initial absolute abundances ($\mathrm{kg\,m^{-3}}$) are defined by the refractories--to--water ice mass ratio $\mu$, the 
molar abundances of CO ($\nu_5$) and $\mathrm{CO_2}$ ($\nu_6$) relative to $\mathrm{H_2O}$, 
and porosity as function of depth. The porosity variation with depth is calculated from hydrostatic equilibrium, as detailed in section~\ref{sec_prep_density_wasteheat}. 
Whether the water ice starts off as amorphous or crystalline, and the partition of CO between pure condensate and $\mathrm{CO_2}$ 
and/or $\mathrm{H_2O}$ hosts, are specified in section~\ref{sec_results} for the individual simulations. When CO is trapped in amorphous $\mathrm{H_2O}$ it is 
here assumed to be fully released upon crystallisation (i.~e., nothing is transferred to the cubic water ice). Note, that if some CO would survive crystallisation by being 
trapped in cubic ice, contrary to this assumption, it would only be released if the temperature reaches 160--$175\,\mathrm{K}$ \citep{barnunetal85}. If cubic ice would be the only 
surviving CO host, it could not possibly explain the observed CO release from comets and Centaurs near and beyond $10\,\mathrm{au}$ \citep{jewitt09} that never reach such temperatures. 
Therefore, the assumed full release of CO upon crystallisation is not a crucial factor in the current discussion. The nominal energy release during crystallisation, and 
its reduction based on the latent heat of released CO are done as in \citet{davidsson21}. 

The appropriate range of $\mu$ for medium--sized TNOs is unknown. The large bodies Pluto and Charon have 
$\mu=0.655\pm0.005$ and $\mu=0.590\pm0.015$, respectively \citep{mckinnonetal17}. \citet{davidssonetal22} demonstrated that Comet 67P 
needs $\mu\approx 1$--2 in order to reproduce the pre-- and post--perihelion water production rate curve. However, other methods have resulted in 
a wide range of estimates for 67P \citep[$0.2\leq\mu<\infty$, as reviewed by][]{choukrounetal20}. I here nominally use $\mu=4$, based on 67P estimates 
early during the \emph{Rosetta} mission \citep{rotundietal15}. Generally, the choice of $\mu$ primarily regulates the level of radiogenic heating within 
the refractory component and heating during crystallisation if the water is amorphous. In the current work, radiogenic heating is ignored (except for a few 
models, as specified). The rather large $\mu$--value therefore primarily acts to keep crystallisation heating relatively low.  This is made intentionally, in order 
to primarily understand the level of processing caused by collisional heating. Secondary model--dependencies on $\mu$ are discussed below.

The applied nominal abundances $\nu_5=0.04$--0.06 (2 per cent as pure ice that is lost before the collisions, and another 2 per cent each in $\mathrm{CO_2}$ and/or $\mathrm{H_2O}$) 
and $\nu_6=0.05$ were based on perihelion coma abundances measured \emph{in situ} at Comet 67P by \citet{hansenetal16}. Those choices were made prior to the demonstration by \citet{davidssonetal22} 
that $\nu_6\approx 0.3$ is needed within the \emph{nucleus} in order to simultaneously fit the observed $\mathrm{H_2O}$ and $\mathrm{CO_2}$ production rate 
curves (including the low perihelion coma $\mathrm{CO_2/H_2O}$ abundance ratio). That $\mathrm{CO_2}$ abundance is close to the average for 50 low--mass 
protostars \citep{pontoppidanetal08}, and if comets have CO abundances similar to such protostars as well, it would suggest $\nu_5=0.13$--$0.26$. In section~\ref{sec_discussion}, I 
discuss the error in $D_{\star}$ (i.~e., the estimated diameter of the largest admissible collisional cascade participant) introduced by ignoring radiogenic heating and 
potentially underestimating the $\mathrm{CO}$ and $\mathrm{CO_2}$ abundances.

\textsc{nimbus} calculates how such an initial setup of abundances and porosities evolves over time due to transport of heat (by solid--state conduction, radiative conduction, 
and advection) and transport of mass (by gas diffusion, driven by sublimation, segregation, crystallisation, and accounting for recondensation processes) in two spatial dimensions (radially and 
latitudinally). The system of coupled differential equations describing such transport, as well as all auxiliary functions, are described by \citet{davidsson21}. 

The temperature--dependent specific heat capacities and heat conductivities of compacted forsterite dust, $\mathrm{H_2O}$, $\mathrm{CO_2}$, and CO used in \textsc{nimbus} 
are all taken from laboratory measurements. Conductivity is corrected for porosity as described by \citet{shoshanyetal02}, assuming $r_{\rm p}=1\,\mathrm{mm}$ pore radii (determining the radiative 
contribution to heat transport, which is negligible at the low temperatures considered here). Laboratory measurements are applied for saturation pressures and latent heats of all volatile species, and 
the equation of state used to calculate porosities under hydrostatic equilibrium. Though these functions must be considered accurate, the bulk specific heat capacity $c(T)$ is somewhat sensitive 
to the assumed $\mu$--value (the water ice specific heat capacity is 11 times higher than that of dust at $50\,\mathrm{K}$, and 4 times higher at $100\,\mathrm{K}$). Changes to 
$\mu$ (for fixed $\mathrm{CO/H_2O}$ and $\mathrm{CO_2/H_2O}$ ratios) also modifies the total mass of $\mathrm{CO}$ and $\mathrm{CO_2}$, thus the amount of energy required to remove or 
relocate those species. The effective solid--state heat conductivity $\kappa_{\rm s}(T,\,\psi)$ is strongly dependent 
on the method used to correct the compacted heat conductivity for porosity $\psi$. The bulk thermal inertia $\Gamma=\sqrt{\rho_{\rm bulk}c(T)\kappa_{\rm s}(T,\,\psi)}$ 
incorporates the combined uncertainties in $c(T)$ and $\kappa_{\rm s}(T,\,\psi)$, and is observationally constrained. The assumptions used in this 
paper result in $90\leq\Gamma\leq 220\,\mathrm{J\,m^{-2}\,K^{-1}\,s^{-1/2}}$ (amorphous water) or $170\leq\Gamma\leq 270\,\mathrm{J\,m^{-2}\,K^{-1}\,s^{-1/2}}$ (crystalline water) at $40\leq T\leq 80\,\mathrm{K}$ (the temperature interval that includes ambient temperature initial conditions at $r_{\rm h}=15$--$30\,\mathrm{au}$, segregation and crystallisation threshold temperatures). The \emph{Philae} lander on Comet 67P has 
performed the only direct \emph{in situ} surface temperature measurements on an outer Solar system minor body, and the resulting initial estimate $\Gamma=85\pm 35\,\mathrm{J\,m^{-2}\,K^{-1}\,s^{-1/2}}$ 
\citep{spohnetal15} has since been revised to $\Gamma\geq 120\,\mathrm{J\,m^{-2}\,K^{-1}\,s^{-1/2}}$ \citep{groussinetal19}. Analysis of remote--sensing irradiation measurements from orbit around 
Comet 67P have resulted in thermal inertia estimates ranging $30\leq\Gamma\leq 160\,\mathrm{J\,m^{-2}\,K^{-1}\,s^{-1/2}}$ \citep[e.~g., ][]{schloerbetal15,marshalletal18,davidssonetal22b}. This suggests that the $\Gamma$--values applied in this work potentially may be on the high side (primarily because the effective heat conductivity may be larger than for at least some real bodies). It should also be kept in mind that the bulk interior 
of medium--sized TNOs may have effective heat conductivities that are different from that of the near--surface region on a single comet nucleus. Tests of how model results change with 
$c(T)$, $\kappa_{\rm s}(T,\,\psi)$, and $\mu$ will be presented in section~\ref{sec_results} and discussed in section~\ref{sec_discussion}.

The gas diffusivity is calculated assuming that pores have lengths $L_{\rm p}=10\,\mathrm{mm}$, radii $r_{\rm p}=1\,\mathrm{mm}$ and unity tortuosity. The gas diffusivity parameters were selected to represent a medium of cm--sized 
porous pebbles, assuming the parent bodies were formed by gravitational pebble--swarm collapse in a streaming instability scenario \citep{youdingoodman05,johansenetal07,nesvornyetal10}. 
Note, that different physically reasonable assumptions about diffusivity will affect the time it takes for vapour to flow from the centre to the surface, but these time--scales are 
so short (years) compared to other physical processes (kyr--Myr) that it has no practical significance for the level of thermal processing. A zero diffusivity (unrealistic for the highly 
porous bodies considered here) would prevent CO from leaving the body altogether, but its internal release and net energy consumption would still take place (i.~e., temperature solutions would not change). 
Because CO could never recondense as pure ice at $r_{\rm h}\leq 30\,\mathrm{au}$, the vapour would remain within the body until the next major collision event, and then be lost to space. $\mathrm{CO_2}$ relocation 
would be inhibited, though.

The current simulations fully account for CO diffusion in the pre--collision simulations that include CO ice. However, for the post--collision simulations where CO is released 
only through segregation and/or crystallisation, immediate escape of CO vapour is assumed. This is because \textsc{nimbus} slows down quite significantly when considering 
CO diffusion with such sources, and this simplification has no effect on the model because it is too warm for CO recondensation. However, full account of $\mathrm{CO_2}$ diffusion is made 
at all times, because recondensation near the cool surface is common. The current simulations assume that the outgassing is so gentle that no dust erosion takes place, i.~e., the body radii 
are not updated over time \citep[for justifications, see][]{davidsson21}.

\section{Preparations} \label{sec_prep}

\subsection{Largest potential parent in a collisional cascade} \label{sec_prep_largest_parent}

The largest parent that may have contributed substantially to the population of $\sim 1\,\mathrm{km}$--sized comets in a collisional cascade must 
fulfil two criteria: 1) it should not be so large and refractory--rich that long--lived radiogenic heating causes $\mathrm{CO_2:CO}$ segregation and 
amorphous $\mathrm{H_2O}$ crystallisation prior to its disruption; 2) the collisional heating itself should not trigger such segregation and crystallisation. 
Unless both criteria are fulfilled, the resulting $\sim 1\,\mathrm{km}$--sized rubble will have little to no CO and be poor comet analogues. 

\begin{figure*}
\centering
\begin{tabular}{cc}
\scalebox{0.4}{\includegraphics{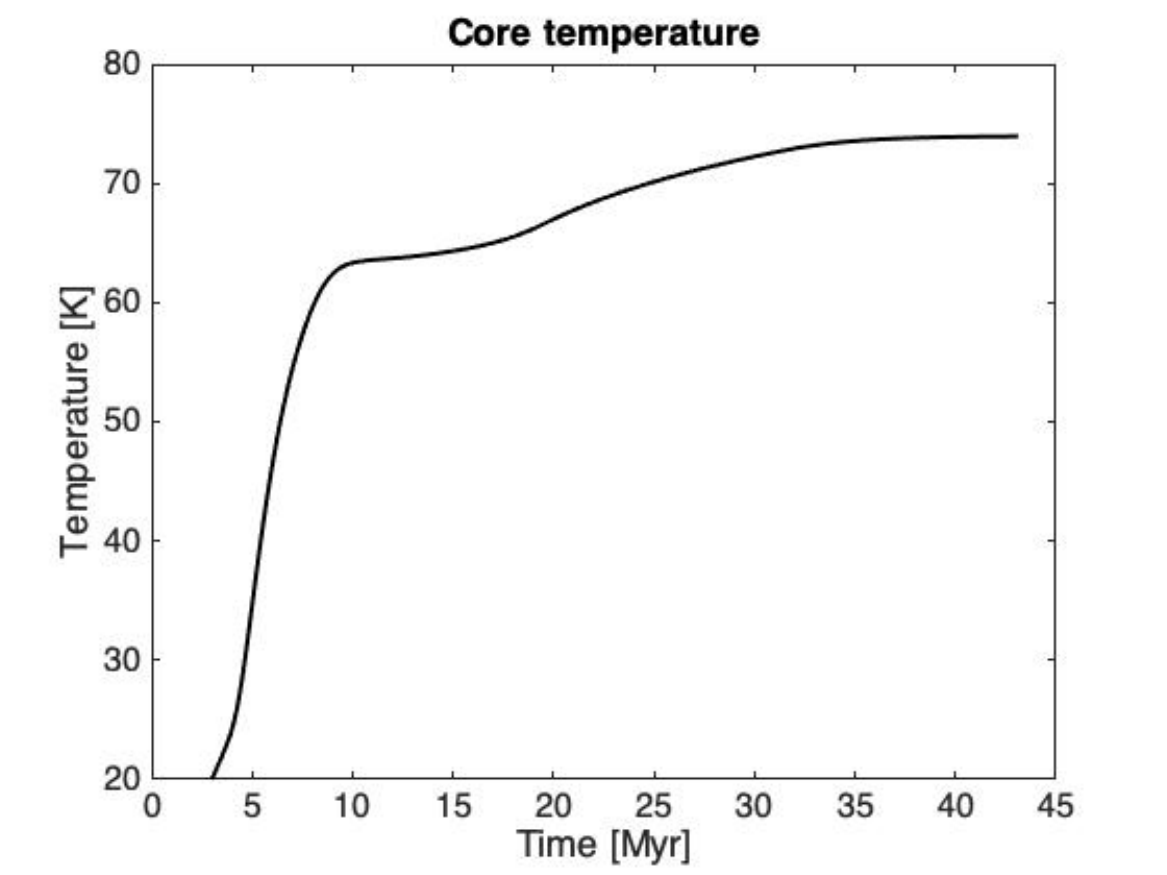}} & \scalebox{0.4}{\includegraphics{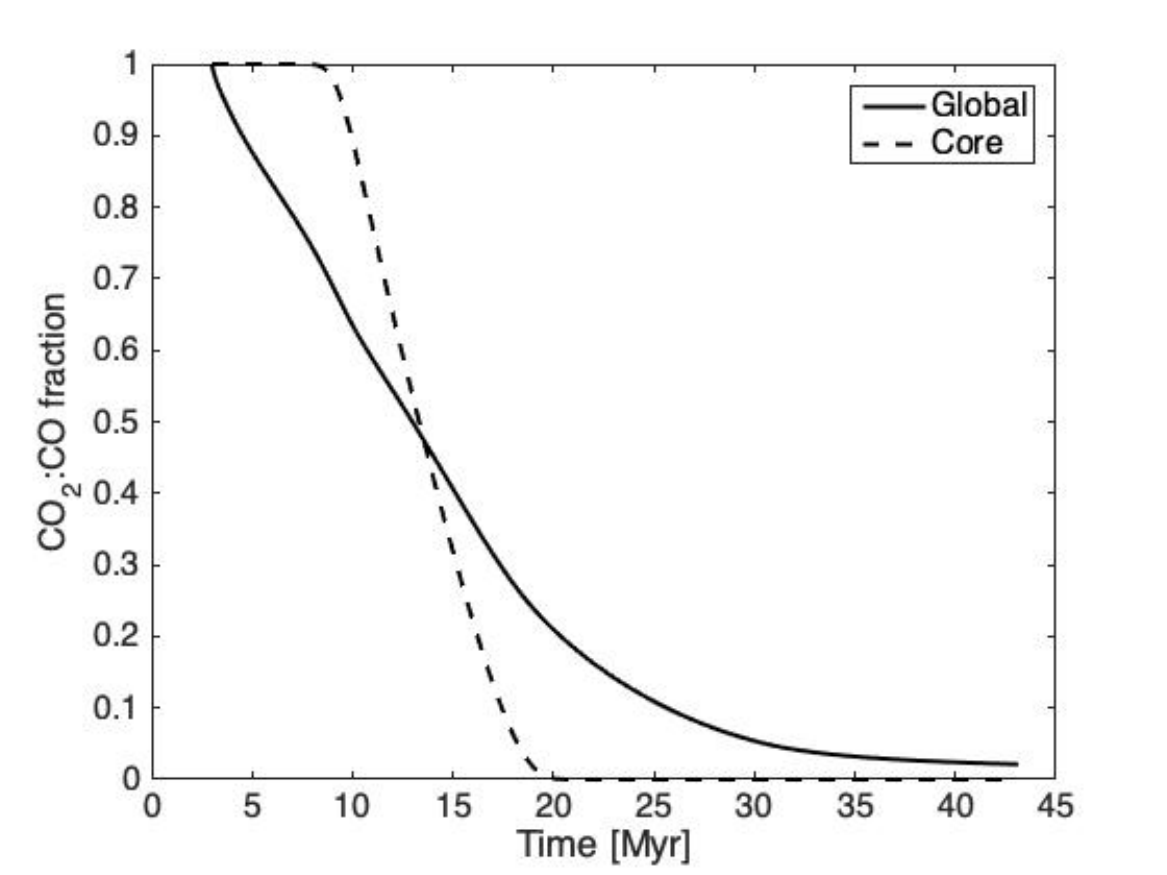}}\\
\scalebox{0.4}{\includegraphics{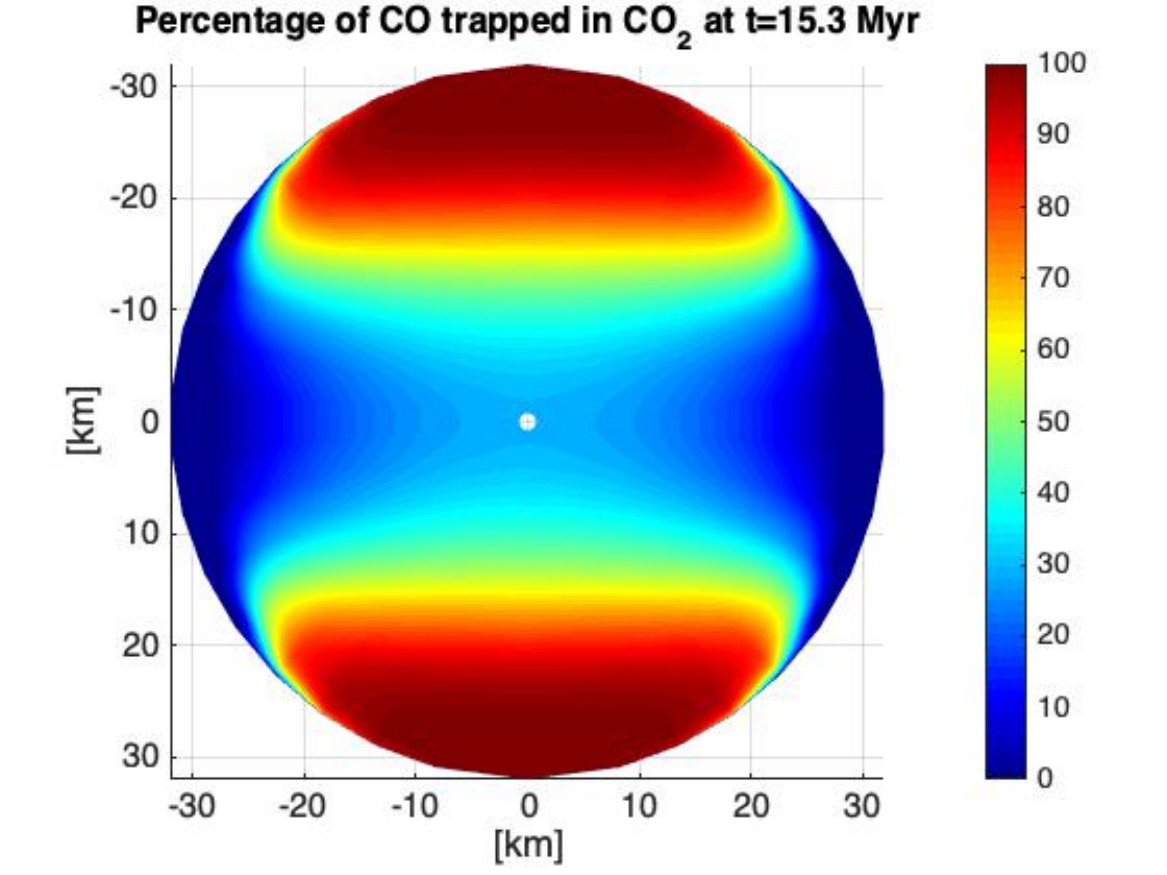}} & \scalebox{0.4}{\includegraphics{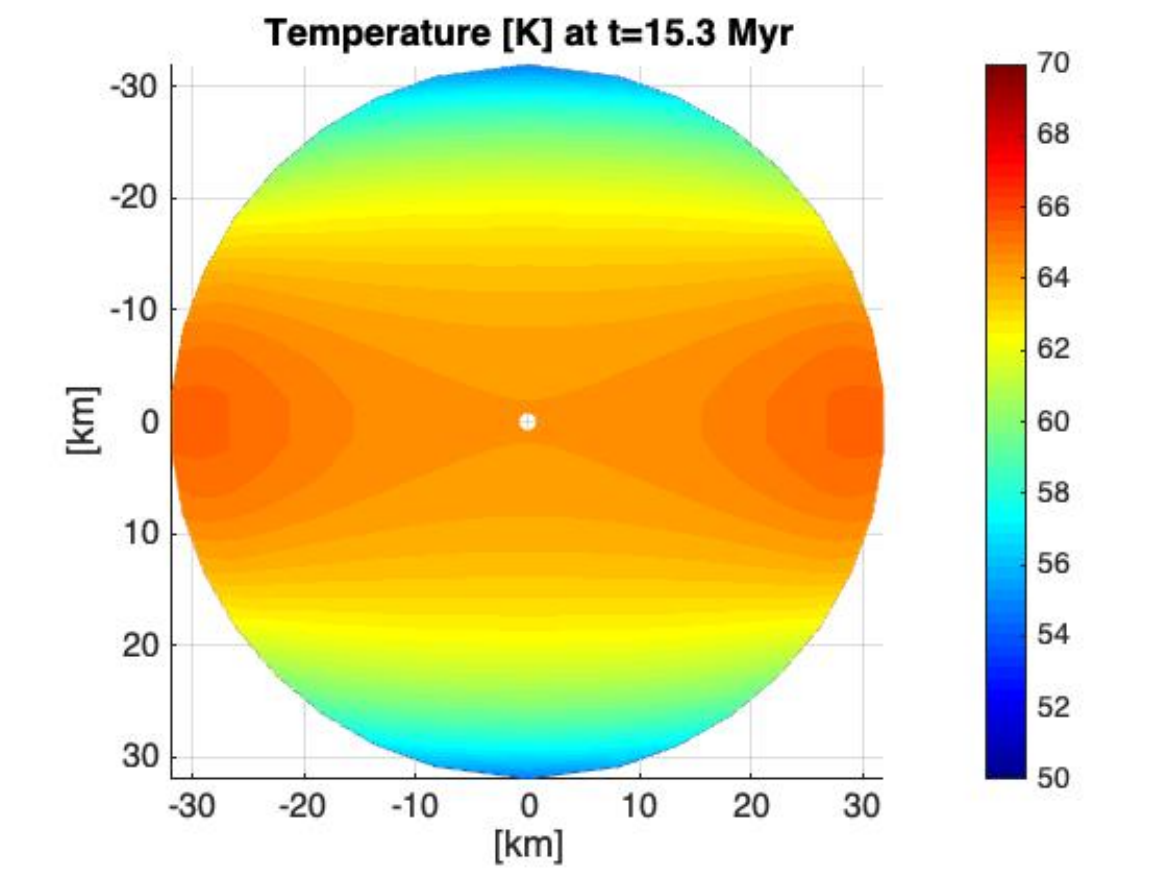}}\\
\end{tabular}
     \caption{A $D=64\,\mathrm{km}$ body at $r_{\rm h}=15\,\mathrm{au}$, consisting of equal masses of amorphous water ice and refractories with chondritic abundances of 
long--lived radionuclides $^{40}\mathrm{K}$, $^{232}\mathrm{Th}$, $^{235}\mathrm{U}$, and $^{238}\mathrm{U}$, with 5 per cent condensed $\mathrm{CO_2}$ and 4 per cent 
CO trapped in equal amounts within the $\mathrm{H_2O}$ and $\mathrm{CO_2}$. The long--term simulation of such a body shows that $\mathrm{CO_2:CO}$ segregation eventually 
completes, but that CO within amorphous $\mathrm{H_2O}$ survives if the body is only subjected to protosolar and radiogenic heating (i.~e., no collisional heating). 
\emph{Upper left:} core temperature as function of time. \emph{Upper right:} the fraction of CO trapped in $\mathrm{CO_2}$ (globally and at the core), versus time. 
\emph{Lower left:} the spatial distribution of CO--bearing $\mathrm{CO_2}$ near the expected catastrophic collision time $t_{10}=15.2\,\mathrm{Myr}$ at this diameter and heliocentric distance (see equation~\ref{eq:05}). 
\emph{Lower right:} the internal temperature distribution at $t_{10}$.}
     \label{fig_D64_test1}
\end{figure*}

\citet{davidsson21} demonstrated that a Hale--Bopp model analogue with $D=74\,\mathrm{km}$, $\mu=4$, and $r_{\rm h}=23\,\mathrm{au}$ first loses all pure $\mathrm{CO}$ ice at $\sim 10\,\mathrm{Myr}$, then 
completes core segregation and crystallisation $\sim 18\,\mathrm{Myr}$ and $\sim 30\,\mathrm{Myr}$ after formation, respectively. 
Unsegregated $\mathrm{CO_2:CO}$ mixtures and amorphous water ice survive in thin near--surface zones where radiative cooling renders 
radiogenic and solar heating insufficient to cause phase transitions. Compared to original abundances, 13 per cent $\mathrm{CO_2:CO}$ and 33 per cent amorphous 
water ice survive. Such deposits would be capable of providing the observed CO outgassing when the body enters the inner Solar system, if the 
body has remained intact since its thermal processing in the primordial disc. However, if such a body would participate in a collisional 
cascade, the resulting rubble would predominantly lack $\mathrm{CO_2:CO}$ mixtures and amorphous water ice. This is because it 
starts out poor in such substances, and presumably would suffer additional losses because of collisional heating. Such rubble would be 
CO--poor, as opposed to most observed comets.

It is therefore likely that the largest participant in a potential collisional cascade in the primordial disc should be smaller and/or contain less 
radionuclide--carrying refractory material than the Hale--Bopp analogue. A suitable body is identified in the following. 

In this paper, it is assumed that Comet 67P with $D_{\rm 67P}=4\,\mathrm{km}$ is the result of a collisional cascade that, in 
each catastrophic collision, reduced the mass of the largest daughter by a factor 2 with respect to the parent. The diameters 
$D_n$ and generation numbers $n$ of the ancestors are related by 
\begin{equation} \label{eq:01}
D_n=2^{n/3}D_{\rm 67P}.
\end{equation}
For example, the immediate parent of Comet 67P (ancestor $n=1$) would be a $D_1=5\,\mathrm{km}$ body, and 
ancestor $n=7$ would be a $D_{7}=20.2\,\mathrm{km}$ body. The largest potential ancestor that still is smaller than the 
Hale--Bopp analogue is $n=12$ with $D_{12}=64\,\mathrm{km}$. I first study how a smaller and less dusty body (than the Hale--Bopp analogue) 
evolves due to heating by long--lived radionuclides (here, $^{40}\mathrm{K}$, $^{232}\mathrm{Th}$, $^{235}\mathrm{U}$, and $^{238}\mathrm{U}$ at chondritic abundances). 
Therefore, a $D=64\,\mathrm{km}$ body with $\mu=1$ (all the water 
initially being amorphous) that contained 5 per cent freely condensed $\mathrm{CO_2}$, and 4 per cent CO (divided equally between 
the $\mathrm{CO_2}$ and amorphous $\mathrm{H_2O}$ hosts) was considered. Note that the CO and $\mathrm{CO_2}$ abundances 
are molar with respect to water. The body was placed on a circular orbit in the ecliptic at $r_{\rm h}=15\,\mathrm{au}$, with a spin axis 
having longitude $\lambda=0^{\circ}$ and latitude $\beta=45^{\circ}$ in the ecliptic system, and was modelled from an assumed 
solar nebula clearing at $t_{\rm c}=3\,\mathrm{Myr}$ to $t=43\,\mathrm{Myr}$. 

Figure~\ref{fig_D64_test1} (upper left) shows the time evolution of the core temperature. The temperature rises from an assumed 
initial temperature of $T_0=20\,\mathrm{K}$ to $T=63\,\mathrm{K}$ at $t=10\,\mathrm{Myr}$. The heating rate 
then slows down because energy is being used to segregate CO out from $\mathrm{CO_2}$. The global and core fractions of 
remaining $\mathrm{CO_2:CO}$ mixtures are shown to the upper right.  Figure~\ref{fig_D64_test1} also shows the internal distributions 
of $\mathrm{CO_2:CO}$ mixture abundance (lower left) and temperature (lower right) and at $t=15.3\,\mathrm{Myr}$. 
That is the point in time when 10 per cent of the original population of $D_{12}=64\,\mathrm{km}$ bodies would have suffered 
catastrophic disruption (see Sec.~\ref{sec_prep_coll_environ}). The $\mathrm{CO_2:CO}$--abundance is heavily depleted except for 
the polar regions (29 per cent of the original $\mathrm{CO_2}$--bound CO remains), and all $\mathrm{CO_2:CO}$ in the core has 
segregated by $t=20.1\,\mathrm{Myr}$. During the first $15\,\mathrm{Myr}$, the hypervolatile CO outgassing is steady at a few times 
$10^{25}\,\mathrm{molec\,s^{-1}}$, while the production rate of the significantly more stable supervolatile $\mathrm{CO_2}$ is 
orders of magnitude lower. 

After the core segregation is completed, the temperature keeps rising above the surface average because of radiogenic heating. 
The core temperature reaches a steady--state value near $74\,\mathrm{K}$ at $t=43\,\mathrm{Myr}$ when the radiative loss rate 
at the surface balances the radiogenic heat production rate. This is merely $\sim 6\,\mathrm{K}$ below the temperature at which 
the Hale--Bopp analogue experienced wide--spread crystallisation. At this time, only 2 per cent of the $\mathrm{CO_2}$--stored CO 
remains, showing that segregation runs to completion at $r_{\rm h}=15\,\mathrm{au}$ in the long run. 

It therefore seems like a body with $D_{12}=64\,\mathrm{km}$ and $\mu=1$ barely would manage to hold on to its most 
resilient CO host, the amorphous water ice, even in the absence of collisional heating. A larger body size, a substantially smaller heat conductivity, 
and/or a higher abundance of refractories ($\mu>1$) would likely lead to eventual large--scale crystallisation.

I therefore settle for $D_{12}=64\,\mathrm{km}$ as the largest parent to be studied in the 
following. Such an ultimate parent in a collisional cascade stands a fair chance of producing CO--rich rubble, particularly if: 1) it is as dust--poor as assumed here; 
2) it experiences the collision relatively early, when radiogenic heat has not had time to accumulate; 3) there are still abundant $\mathrm{CO_2:CO}$ mixtures that 
can absorb collisional energy during segregation and help preserving CO stored in amorphous water ice. However, larger bodies could not 
avoid losing all their $\mathrm{CO}$ through a combination of radiogenic and collisional heating (unless being so massive that self--gravity prevents escape). They are prevented from participating in a 
collisional cascade, because they would produce a population of $\mathrm{CO}$--free comet--sized bodies that are not present in the 
observational record.

\subsection{Collisional environment} \label{sec_prep_coll_environ}

In order to describe the collisional environment, I consider the primordial disc at heliocentric distances $15\leq r_{\rm h}\leq 30\,\mathrm{au}$. 
It contains bodies with diameter $0.02\leq D\leq 1000\,\mathrm{km}$, having a differential size--frequency distribution power--law index $q=-3$, i.~e., 
the number of bodies with diameter $D$ is $\mathcal{N}(D)\propto D^q$. It is assumed that the primordial disc contains $N_{D>D_{\rm lim}}=2\cdot 10^{11}$ bodies with diameters $D_{\rm lim}\geq 2.3\,\mathrm{km}$ 
\citep{brassermorbidelli13,morbidellirickman15,rickmanetal15}. The surface density is assumed to go as $\propto r_{\rm h}^{-1}$. The primordial disc is divided into 
Zone~\#1 ($15\leq r_{\rm h}< 20\,\mathrm{au}$), Zone~\#2 ($20\leq r_{\rm h}< 25\,\mathrm{au}$), and Zone~\#3 ($25\leq r_{\rm h}\leq 30\,\mathrm{au}$), 
each containing a fraction $f_{\rm z}=1/3$ of the bodies, consistent with the considered surface density. The number of impacts $\mathrm{[yr^{-1}]}$ on a target body of diameter $D_{\rm p}$ 
within Zone~\#$i$, by projectiles with diameters $d_{\rm min}\leq d_{\rm proj}\leq d_{\rm max}$ originating from Zone~\#$j$ is given by
\begin{equation} \label{eq:02}
C_{ij}=\frac{1}{4}P_{ij}\int_{d_{\rm min}}^{d_{\rm max}}\left(D_{\rm p}+d_{\rm proj}\right)^2f_{\rm z}\mathcal{N}(d_{\rm proj})\,dd_{\rm proj}.
\end{equation}

Here, $P_{ij}$ is the mean intrinsic collision probability for targets in Zone~$\#i$ and projectiles originating from Zone~$\#j$, with numerical 
values from \citet{morbidellirickman15} in Table~\ref{tab_morbrick}. For targets in Zone~\#1, 83 per cent of the projectiles originate from within 
the same zone. For Zone~\#3, the corresponding number is 90 per cent. These fractions are rather high, because they each have a single neighbouring zone. 
For Zone~\#2 targets, 66 per cent of the projectile come from the same zone, while 28 per cent come from Zone~\#1 and 5 per cent from Zone~\#3. For simplicity, 
only same--zone targets and projectiles are considered here (i.~e., $i=j$), in order to assign a single typical projectile size and velocity that causes a catastrophic 
disruption of a given target body in a given zone. However, it should be remembered that the real lifetimes are somewhat shorter than calculated here, 
and that some targets will be destroyed by unusually small (and fast) projectiles originating from other zones.

\begin{table}
\begin{center}
\begin{tabular}{||l|r|r|r||}
\hline
\hline
Parameter & Zone~\#1 & Zone~\#2 & Zone~\#3\\
\hline
$P_{ij}\,\mathrm{[yr^{-1}\,proj^{-1}\,targ^{-1}]}$ & $1.85\cdot 10^{-20}$ & $8.95\cdot 10^{-21}$ & $7.32\cdot 10^{-21}$\\
$V_{ij}\,\mathrm{[m\,s^{-1}]}$ & 780 & 440 & 240\\
\hline
\end{tabular}
\caption{Mean intrinsic collision probability $P_{ij}$ and mean collision velocity $V_{ij}$ for targets and projectiles in the same zones, 
taken from \citet{morbidellirickman15}.}
\label{tab_morbrick}
\end{center}
\end{table}

In equation~\ref{eq:02}, $d_{\rm min}$ is the diameter of the smallest projectile capable of causing a catastrophic collision. It is calculated by first considering 
the critical specific catastrophic collisional energy $Q_{\rm D}^*\,\mathrm{(J\,kg^{-1})}$, for which the largest surviving fragment after the collision carries half the mass of the parent,
\begin{equation} \label{eq:03}
Q_{\rm D}^*=a_{\rm coll}\left(D_{\rm p}/2\right)^{3\mu_{\rm coll}}V_{ij}^{2-3\mu_{\rm coll}}
\end{equation}
with $a_{\rm coll}=4\cdot 10^{-4}$ and coupling parameter $\mu_{\rm coll}=0.42$ for weak, highly porous bodies according to \citet{jutzietal17}, 
and with impact velocity $V_{ij}$ from \citet{morbidellirickman15} given in Table~\ref{tab_morbrick}. By equating the total energy needed for 
the catastrophic disruption of the target with the kinetic energy of the projectile, one can solve for $d_{\rm min}$ (assuming the two bodies have the same density),
\begin{equation} \label{eq:04}
d_{\rm min}=\frac{(2Q_{\rm D}^*)^{1/3}D_{\rm p}}{V_{ij}^{2/3}}.
\end{equation}
In order to define an upper limit for the integral in equation~(\ref{eq:02}), projectiles carrying $2Q_{\rm D}^*$ are considered.

The probability that an object of diameter $D_{\rm p}$ is intact after a time $t$ is given by $p_{\rm int}=\exp(-C_{ij}t)$ according to \citet{morbidellirickman15}. In order to define 
a timescale for substantial destruction of $D_{\rm p}$--type parents, I apply $p_{\rm int}=0.9$ and
\begin{equation} \label{eq:05}
t_{10}=-\frac{\ln(0.9)}{C_{ij}},
\end{equation}
i.~e., $t_{10}$ is the time it would take to destroy 10 per cent of the $D_{\rm p}$ population. This time--scale is admittedly arbitrary, but is meant to measure the time it 
would take for bodies of a given size to have started to contribute significantly to the catastrophic cascade. Solutions to equations~(\ref{eq:01})--(\ref{eq:02}) for Zones~\#1--\#3 
are shown in Tables~\ref{tab_z1a}--\ref{tab_z3a}, respectively. For example, a $D_{\rm p}=64\,\mathrm{km}$ body in Zone~\#2 would wait for $t_{10}=44.2\,\mathrm{Myr}$ to get destroyed by a $d_{\rm proj}=38.2\,\mathrm{km}$ 
projectile in a collision with $Q_{\rm D}^*=1.72\cdot 10^4\,\mathrm{J\,kg^{-1}}$ that would create a $D_{\rm d}=50.8\,\mathrm{km}$ daughter. 

\begin{table*}
\begin{center}
\begin{tabular}{||r|r|r|r|r|r|r|r|r|r||}
\hline
\hline
$n$ & $D_{\rm p}\,\mathrm{[km]}$ & $t_{10}\,\mathrm{[Myr]}$ & $d_{\rm proj}\,\mathrm{[km]}$ & $Q_{\rm D}^*\,\mathrm{[kJ\,kg^{-1}]}$ & $D_{\rm d}\,\mathrm{[km]}$ & $\rho_{\rm block}\,\mathrm{[kg\,m^{-3}]}$ & $\psi_{\rm macro}$ & $\rho_{\rm bulk}\,\mathrm{[kg\,m^{-3}]}$ & $f_{\rm waste}$\\
\hline
12  & 64.0  & 15.2 & 26.0 & 26.2 & {\bf  50.8} &  1300 & 0.57 & 556 & 0.68\\
11  & 50.8 & 13.4 & 18.7 & 19.6 & {\bf  40.3} &  1300 & 0.58 & 548 & 0.74\\
10 & 40.3 & 11.6 & 13.5 & 14.6 & {\bf 32.0} & 1300 & 0.58 & 544 & 0.83\\
9 & 32.0 & 10.2 & 9.7 & 10.9 &  {\bf 25.4} & 1300 & 0.58 & 541 & 0.94\\
8 & 25.4 & 8.8 & 7.0 & 8.18 &  {\bf 20.2} & 1264 & 0.57 & 539 & 1.00\\
7 & 20.2 & 7.5 & 5.1 & 6.13 & {\bf 16.0} & 1219 & 0.56 & 538 & 1.00\\
\hline
1 & 5.0 & 2.8 & 0.7 & 1.06 & {\bf 4.0} & 1033 & 0.48 & 537 & 1.00\\
\hline
\end{tabular}
\caption{Zone~\#1 at $15$--$20\,\mathrm{au}$: the $n^{\rm th}$ generation parent of 67P with diameter $D_{\rm p}$ waits a time $t_{10}$ for a projectile of 
diameter $d_{\rm proj}$ to impact with a specific energy $Q_{\rm D}^*$, thereby creating a daughter with diameter $D_{\rm d}$. The parent material has compressed to blocks of 
density $\rho_{\rm block}$ that reassemble with macro porosity $\psi_{\rm macro}$ to give the daughter a bulk density $\rho_{\rm bulk}$. A fraction $f_{\rm waste}$ of $Q_{\rm D}^*$ 
goes into heating the daughter.}
\label{tab_z1a}
\end{center}
\end{table*}

\subsection{Bulk density and waste heat} \label{sec_prep_density_wasteheat}

The next step is to calculate pre--collision bulk densities and bulk porosities for the targets. That is necessary in order to evaluate the 
expected degree of compaction taking place during catastrophic collisions, which in turn determines what fraction of $Q_{\rm D}^*$ 
that should go into waste heat. To this end, the hydrostatic equilibrium configurations of the parent bodies are calculated, for which 
the sum of gravitational pressure and dynamic pressure due to accretion, at any given depth, is balanced by the compressive strength of the material. 
The gravitational pressure is given by \citep[e.g.][]{henkeetal12}
\begin{equation} \label{eq:06}
\mathcal{P}_{\rm g}(r)=-4\pi G\int_r^R\frac{\rho(r')}{(r')^2}\left(\int_0^r\rho(r') (r')^2\,dr'\right)\,dr',
\end{equation}
and the dynamic pressure is given by
\begin{equation} \label{eq:07}
\mathcal{P}_{\rm d}=\frac{1}{2}\rho_{\rm imp}V_{\rm imp}^2
\end{equation}
where $\rho_{\rm imp}$ and $V_{\rm imp}$ are the mean density and velocity of the material that came together to 
form the body during primordial accretion. I require that these compressive pressures should be balanced ($\mathcal{P}_{\rm g}+\mathcal{P}_{\rm d}=\mathcal{P}_{\rm m}$) 
by the compressive strength $\mathcal{P}_{\rm m}$, at which a granular material manages to resists further compression below the local porosity $\psi=\psi (r)$. 
The function $\mathcal{P}_{\rm m}=\mathcal{P}_{\rm m}(\psi)$ is calculated as a weighted average of the 
compressive strengths measured for silica particles (representing refractories) and water ice. For refractories, the omnidirectional version of equation~(10) in  \citet{guettlereta09} is applied. 
For ice, the measurements by \citet{loreketal16} are applied for pressures $<10^5\,\mathrm{Pa}$, but above this threshold I use data from \citet{yasuiarakawa09} obtained 
for a $\mathrm{ref:ice}=29:71$ mixture at $T=206\,\mathrm{K}$ (see their Fig.~2). I assume densities $\rho_{\rm ref}=3000\,\mathrm{kg\,m^{-3}}$ and $\rho_{\rm ice}=960\,\mathrm{kg\,m^{-3}}$ 
for refractories and ice mixtures, respectively. The mass fractions are taken as $f_{\rm ref}=0.53$ and $f_{\rm ice}=0.47$, which implies a volumetric fraction of ices $f_{\rm V,ice}=0.63$ that 
is used as a weighting factor for the compressive strengths.

The compact density of this mixture of refractories and ices is $\rho_{\rm solid}=1300\,\mathrm{kg\,m^{-3}}$. In order to evaluate $\mathcal{P}_{\rm d}$ I assume $\rho_{\rm imp}=300\,\mathrm{kg\,m^{-3}}$ 
and I require that a cloud with the mass of Comet 67P should compress into a $D_{\rm 67P}=4\,\mathrm{km}$ body with the bulk density near $\rho_{\rm bulk}=535\,\mathrm{kg\,m^{-3}}$ 
\citep{jordaetal16,preuskeretal15,patzoldetal19}. The hydrostatic equilibrium calculations show that this happens if $V_{\rm imp}=31\,\mathrm{m\,s^{-1}}$. The same $\mathcal{P}_{\rm d}$--value is 
applied in all calculations. Tables~\ref{tab_z1a}--\ref{tab_z3a} show the resulting bulk densities $\rho_{\rm bulk}$ as function of body size. Self--gravity is rather weak at these body sizes, e.~g., 
the $D_{\rm p}=64\,\mathrm{km}$ body compresses to $\rho_{\rm bulk}=556\,\mathrm{kg\,m^{-3}}$, marginally higher than that of Comet 67P. The porosity grows from $\psi=0.55$ at the centre to 
$\psi=0.58$ near the surface, giving a mean porosity of $\psi=0.57$ (compared to the modelled value $\psi_{\rm 67P}=0.59$ for Comet 67P), and a rather homogeneous interior (this level 
of variation is too small to have any practical influence).

I now calculate the work needed to compress one kilogram of material from the porosity $\psi_{67P}=0.59$ and bulk density $\rho_{\rm bulk,67P}=537\,\mathrm{kg\,m^{-3}}$ 
(occupying a volume $V_1=\rho_{\rm bulk,67P}^{-1}$) of Comet 67P, to full compaction (here, $\psi_{\rm min}=0.006$, $\rho_{\rm max}=(1-\psi_{\rm min})\rho_{\rm solid}=1292\,\mathrm{kg\,m^{-3}}$, 
occupying a volume $V_0=\rho_{\rm max}^{-1}$), using the compression strength $\mathcal{P}_{\rm m}=\mathcal{P}_{\rm m}(\psi)$ discussed above,
\begin{equation} \label{eq:08}
Q_{\rm max}=-\int_{V_0}^{V_1}\mathcal{P}_{\rm m}(\psi)\,dV.
\end{equation}
Doing so, it is found that the largest amount of waste heat one can expect from porosity--removal alone is $Q_{\rm max}=9.59\,\mathrm{kJ\,kg^{-1}}$. 
According to \citet{benavidez_campobagatin09}, the fraction of $Q_{\rm D}^*$ partitioned to kinetic energy of the ejecta is
\begin{equation} \label{eq:08b}
f_{\rm KE}=\frac{k}{k-2}2^{-2/k}\frac{V_{\rm esc}^2}{Q_{\rm D}^*}
\end{equation}
where $k=9/4$ and $V_{\rm esc}$ is the escape velocity. It is so small ($f_{\rm KE}<0.04$) that this loss is ignored in the following. 
In case $Q_{\rm D}^*\leq Q_{\rm max}$ it is here assumed that all available energy goes into compaction and consequently is transformed to 
waste heat (the fraction $f_{\rm waste}$ of $Q_{\rm D}^*$ being waste heat is unity). Table~\ref{tab_z1a} shows that this happens for disruption 
of parents with $D_{\rm p}\leq 25.4\,\mathrm{km}$ at $r_{\rm h}=15\,\mathrm{au}$, and Table~\ref{tab_z3a} shows that the limit is changed to 
$D_{\rm p}\leq 50.8\,\mathrm{km}$ at $r_{\rm h}=30\,\mathrm{au}$.

The integral in equation~(\ref{eq:08}) can be used to evaluate the volume of a $1\,\mathrm{kg}$ mass 
element resulting from using an energy $Q_{\rm D}^*$ for compression. The corresponding density $\rho_{\rm block}$ of the partially compressed fragments 
that will build the daughter bodies are listed in Tables~\ref{tab_z1a}--\ref{tab_z3a} as well.  When $\rho_{\rm block}<\rho_{\rm solid}$ the blocks will have some 
level of micro--porosity. For simplicity, it is here assumed that the daughters built by gravitational reaccumulation of the partially compressed blocks with 
$\rho_{\rm block}$ will have bulk densities that are equivalent to the ones caused by primordial formation (i.~e., $\rho_{\rm bulk}$).  That explains 
the values for macro--porosity $\psi_{\rm macro}$ for the daughter bodies in Tables~\ref{tab_z1a}--\ref{tab_z3a}. According to this formalism, the bulk density of 
blocks within Comet 67P, had it been formed as the result of a catastrophic disruption of a $D_{\rm p}=5\,\mathrm{km}$ parent at 20--$25\,\mathrm{au}$, would have 
been $\rho_{\rm block}\approx 990\,\mathrm{kg\,m^{-3}}$.

In case $Q_{\rm D}^*>Q_{\rm max}$, the parent body material is assumed to reach full compression, $\rho_{\rm block}=\rho_{\rm solid}$. For a brief 
moment, the material will be shocked to densities higher than $\rho_{\rm solid}$, and it is here assumed that the work performed to compress 
above $\rho_{\rm solid}$ is twice as large as that performed by the release wave when expanding the material back to $\rho_{\rm solid}$. This estimate is 
based on measurements of the shock and particle velocities in water ice IV by \citet{stewartahrens05}, applied to the method of calculating waste heat by \citet{sharpdecarli06}, 
indicating that about half of the internal energy increase is converted to waste heat in water ice at impact pressures $\leq 3\,\mathrm{GPa}$. Consequently, the total waste heat is taken as
\begin{equation} \label{eq:09}
Q_{\rm waste}=\frac{1}{2}\left(Q_{\rm D}^*-Q_{\rm max}\right)+Q_{\rm max}.
\end{equation}
The corresponding $f_{\rm waste}=Q_{\rm waste}/Q_{\rm D}^*$ values are listed in Tables~\ref{tab_z1a}--\ref{tab_z3a}. Note that daughters formed from, e.~g., 
$D_{\rm p}\geq 32\,\mathrm{km}$ at $r_{\rm h}=15\,\mathrm{au}$ parents therefore have $\rho_{\rm block}=\rho_{\rm solid}=1300\,\mathrm{kg\,m^{-3}}$. As previously 
mentioned, it is assumed that the reaccumulated daughters of such parents have bulk densities identical to the $\rho_{\rm bulk}$--values expected from primordial formation 
(the $\psi_{\rm macro}$--values are calculated accordingly). For further discussion of waste heat during compression, see Appendix~\ref{appendix01}.

\begin{table*}
\begin{center}
\begin{tabular}{||r|r|r|r|r|r|r|r|r|r||}
\hline
\hline
$n$ & $D_{\rm p}\,\mathrm{[km]}$ & $t_{10}\,\mathrm{[Myr]}$ & $d_{\rm proj}\,\mathrm{[km]}$ & $Q_{\rm D}^*\,\mathrm{[kJ\,kg^{-1}]}$ & $D_{\rm d}\,\mathrm{[km]}$ & $\rho_{\rm block}\,\mathrm{[kg\,m^{-3}]}$ & $\psi_{\rm macro}$ & $\rho_{\rm bulk}\,\mathrm{[kg\,m^{-3}]}$ & $f_{\rm waste}$\\
\hline
12 & 64.0 & 44.2 & 38.2 &  17.2 & {\bf 50.8} & 1300  & 0.57  & 556  & 0.78\\
11 & 50.8  & 39.3  &27.5 &  12.8 & {\bf 40.3}  & 1300  & 0.58  & 548  &  0.87 \\
10 & 40.3 & 34.6 & 19.8 &  9.59 & {\bf 32.0} & 1292  & 0.58 & 544 & 1.00\\
9 & 32.0 & 30.3 & 14.3 &  7.17 & {\bf 25.4} &  1243  & 0.56  & 541  &  1.00\\
8 & 25.4 & 26.7 & 10.3 &  5.36 & {\bf 20.2} & 1201  & 0.55 & 539 & 1.00\\
7 & 20.2 &  23.1 & 7.4 &  4.01 & {\bf 16.0} &  1166 & 0.54 & 538 &  1.00\\
\hline
1 & 5.0 &  9.8 & 1.0 & 0.69 & {\bf 4.0} & 987  & 0.46 &  537  &  1.00\\
\hline
\end{tabular}
\caption{Zone~\#2 at $20$--$25\,\mathrm{au}$: the $n^{\rm th}$ generation parent of 67P with diameter $D_{\rm p}$ waits a time $t_{10}$ for a projectile of 
diameter $d_{\rm proj}$ to impact with a specific energy $Q_{\rm D}^*$, thereby creating a daughter with diameter $D_{\rm d}$. The parent material has compressed to blocks of 
density $\rho_{\rm block}$ that reassemble with macro porosity $\psi_{\rm macro}$ to give the daughter a bulk density $\rho_{\rm bulk}$. A fraction $f_{\rm waste}$ of $Q_{\rm D}^*$ 
goes into heating the daughter.}
\label{tab_z2a}
\end{center}
\end{table*}

\begin{table*}
\begin{center}
\begin{tabular}{||r|r|r|r|r|r|r|r|r|r||}
\hline
\hline
$n$ & $D_{\rm p}\,\mathrm{[km]}$ & $t_{10}\,\mathrm{[Myr]}$ & $d_{\rm proj}\,\mathrm{[km]}$ & $Q_{\rm D}^*\,\mathrm{[kJ\,kg^{-1}]}$ & $D_{\rm d}\,\mathrm{[km]}$ & $\rho_{\rm block}\,\mathrm{[kg\,m^{-3}]}$ & $\psi_{\rm macro}$ & $\rho_{\rm bulk}\,\mathrm{[kg\,m^{-3}]}$ & $f_{\rm waste}$\\
\hline
12 & 64.0 & 70.8 & 46.4 & 11.0 & {\bf  50.8} & 1300 & 0.57 & 556 & 0.94\\
11 & 50.8 & 63.4 & 33.4 & 8.19 & {\bf 40.3} & 1264 & 0.57 & 548 & 1.00\\
10 & 40.3 & 56.5 & 24.0 & 6.12 & {\bf 32.0} & 1219 & 0.55 & 544 & 1.00\\
9 & 32.0 & 50.3 & 17.3 & 4.58 & {\bf 25.4} & 1181 & 0.54 & 541 & 1.00\\
8 & 25.4 & 44.5 & 12.5 & 3.42 & {\bf 20.2} & 1148 & 0.53 & 539 & 1.00\\
7 & 20.2 & 38.9 & 9.0 & 2.56 & {\bf 16.0} & 1119 & 0.52 & 538 & 1.00\\
\hline
1 & 5.0 & 17.0 & 1.2 & 0.44 & {\bf 4.0} & 933 & 0.42 & 537 & 1.00\\
\hline
\end{tabular}
\caption{Zone~\#3 at $25$--$30\,\mathrm{au}$: the $n^{\rm th}$ generation parent of 67P with diameter $D_{\rm p}$ waits a time $t_{10}$ for a projectile of 
diameter $d_{\rm proj}$ to impact with a specific energy $Q_{\rm D}^*$, thereby creating a daughter with diameter $D_{\rm d}$. The parent material has compressed to blocks of 
density $\rho_{\rm block}$ that reassemble with macro porosity $\psi_{\rm macro}$ to give the daughter a bulk density $\rho_{\rm bulk}$. A fraction $f_{\rm waste}$ of $Q_{\rm D}^*$ 
goes into heating the daughter.}
\label{tab_z3a}
\end{center}
\end{table*}

\subsection{Methodology} \label{sec_prep_method}

The degree of thermal processing during a catastrophic impact event depends on the initial composition of 
the body and on the thermal evolution that took place prior to the collision. If condensed CO remains at the time of 
impact, it can act as a heat sink that protects potential deposits of CO within $\mathrm{CO_2}$ and/or amorphous 
$\mathrm{H_2O}$. Similarly, segregation can delay or prevent the onset of $\mathrm{CO_2}$ redistribution or loss, and $\mathrm{H_2O}$ crystallisation. 
Therefore, a number of pre--collision simulations are performed with the purpose of: 1) finding the loss time--scale of freely 
condensed CO; 2) to investigating the stability of $\mathrm{CO_2:CO}$ mixtures; 3) investigating the stability of $\mathrm{CO_2}$ ice.
These models consider $7\leq n\leq 12$ parents with diameters in the range $20.2\leq D\leq 64\,\mathrm{km}$. The lower cut--off is 
placed at sizes for which substantial volatile loss is not expected, to be verified during the simulations. These bodies are all considered to 
be the first parent (i.~e., the largest body) participating in the collisional cascade, i.~e., they are not themselves formed as the result of 
a catastrophic collision. This is done in order to attempt to define the approximate starting--point of a potential collisional cascade, i.~e., 
the largest body size that substantially may have contributed to the current population of $\sim 1\,\mathrm{km}$--class cometary collisional rubble. 
Furthermore, potential radiogenic heating by short--lived radionuclides ($^{26}$Al and $^{60}$Fe) is ignored, assuming that those effects in any case 
must have been small enough to leave the $\mathrm{CO_2:CO}$ mixtures and amorphous $\mathrm{H_2O}$ intact. 

Ideally, such models should be run from solar nebula clearing (here taken as $t_{\rm c}=3\,\mathrm{Myr}$) to the time of impact at $t_{10}$. 
However, performing simulations for such long time intervals is computationally prohibitive due to the large number of cases considered 
in this paper. Therefore, a first set of models are run up to the point when free CO ice is lost. Some models are continued further at various 
degrees, in order to explore the stability of $\mathrm{CO_2:CO}$ mixtures and $\mathrm{CO_2}$ ice and to estimate their loss rates. 
However, then it is necessary to jump ahead in time. 

Therefore, a second set of models is then initiated briefly before $t_{10}$. They consider the post--collision daughter nuclei to the previously studied 
parents, i.~e., focus on $6\leq n\leq 11$ bodies with diameters in the range $16\leq D\leq 50.8\,\mathrm{km}$. These are run for a period prior 
to impact to allow for global thermal relaxation to the protosolar luminosity conditions prevailing at the time. At $t_{10}$, an amount of 
energy $Q_{\rm D}^*f_{\rm waste}$ is injected homogeneously, in order to simulate a newly formed rubble pile, consisting of heated debris from the parent. 
\citet{leinhardtandstewart09} demonstrated that the interior of collisional rubble piles consists of moderately shocked (thus moderately heated) material, while 
the surface is a mixture of weakly and strongly shocked (thus coldest or warmest) material. The assumption of homogeneous energy injection is therefore 
equivalent to assuming that the internal energy density carried by the coldest or warmest material near the surface averages out to a level similar to the 
internal energy of core material. For technical reasons, this energy is injected gradually during a $100\,\mathrm{yr}$ period, which is very short compared to the 
post--collision cooling timescale back to ambient temperature conditions. Changes to the abundances of $\mathrm{CO_2:CO}$ mixtures, amorphous $\mathrm{H_2O}$ ice, 
and $\mathrm{CO_2}$ is are monitored during cooling. 

Skipping millions of years of evolution between the two sets of simulations means that it is not possible to account for heating by 
long--lived radionuclides. Comparisons between the model in Sec.~\ref{sec_prep_largest_parent} with similar simulations performed 
without radiogenic heating show that the temperature elevation caused by radioactive decay is $\sim 10\,\mathrm{K}$ for $D=64\,\mathrm{km}$ 
bodies. The discrepancy falls rapidly with decreasing body diameter. As argued in section~\ref{sec_discussion}, this effect could be partially or 
fully cancelled if the $\mathrm{CO}$ abundance is higher than currently assumed.

The post--collision simulations are treated differently in the three zones. At $15\leq r_{\rm h}<20\,\mathrm{au}$ (Zone\#1) it is necessary to include 
amorphous water ice, because $\mathrm{CO_2:CO}$ mixtures are unstable at such distances, and amorphous $\mathrm{H_2O}$ provides 
the only stable refuge for hypervolatiles in the absence of collisions.  The effect of collisional heating on this reservoir needs to be investigated, 
both with and without the heat sink provided by residual $\mathrm{CO_2:CO}$, potentially being present at the time of collision. However, we 
do not know for certain that amorphous water ice exists. If it does not exist, we may have to accept that: 1) small objects that originated from Zone\#1 no 
longer carries hypervolatiles; 2) $\mathrm{CO_2}$ is the sole current reservoir of hypervolatiles in comets. At $20<r_{\rm h}<25\,\mathrm{au}$ (Zone\#2) 
I assume this is the case, i.~e.,  the stability of $\mathrm{CO_2:CO}$ against collisional processing is investigated, without the inclusion of 
amorphous ice. One reason for doing this is to study the effect of collisions on the mobilisation and internal redistribution of $\mathrm{CO_2}$ ice, 
without the energy release associated with crystallisation. At $25<r_{\rm h}\leq 30\,\mathrm{au}$ (Zone\#3), amorphous water ice is included anew (in order 
to test its resilience against collisional heating both a small and large distances). 

In the following, the results of numerous simulations are presented. The model bodies are designed to obey specific 
physical relations (e.~g., the saturation pressure of $\mathrm{CO_2}$ vapour as function of temperature), and are assigned specific values of the 
free model parameters (e.~g., the length, width, and tortuosity of tubes that define gas diffusivity). As far as possible, all necessary physical 
relations are based on laboratory measurements of materials believed to be common in minor outer Solar System bodies. Free parameter values are 
educated guesses, based on theoretical predictions and fits to observed behaviour. A model body of a given size is thereby designed to behave 
in one specific way when subjected to given internal and external stimuli (collisional and solar energy). However, the real Solar System bodies of 
that size most likely have a range of properties depending on individual formation and evolution conditions. Thus, they are not identical to one 
another and will behave differently when subjected to the same stimuli. It will be assumed that the model bodies behave quantitatively similar 
to a sub--class of real bodies with the same size. The model simulations will be used to determine whether or not a real body of a given size 
experiences sufficiently small abundance changes in a catastrophic disruption to be considered a suitable comet nucleus ancestor. From 
the discussion above it is clear that such statements may only be correct for a fraction of the bodies of a given size. This uncertainty should be kept in 
mind when a limit $D_{\star}$ is placed on the largest acceptable ancestor in a collisional cascade -- it is a best--effort assessment that 
problems of keeping CO have started for bodies twice as massive, 
but it is impossible to certify whether those problems are limited or affects the entire size--class. All that can be said is 
that problems ought to be significantly smaller for most bodies having half the mass of a $D_{\star}$ body, and significantly larger 
for most bodies with twice its mass and above.

\section{Results} \label{sec_results}

\subsection{Inner disc thermal evolution} \label{sec_results_inner}

\subsubsection{Pre--collision evolution at $r_{\rm h}=15\,\mathrm{au}$} \label{sec_results_inner_pre}

For the pre--collision models, it is assumed that newly formed planetesimals have $\mu=4$, a $\mathrm{CO_2}$ molar abundance of 
5 per cent relative water, and a 4 per cent molar abundance of CO relative water that is evenly distributed between freely condensed CO and 
CO entrapment within $\mathrm{CO_2}$. As described in Sec.~\ref{sec_prep_method}, these simulations are performed without radiogenic heating. 
Additionally, the water ice is here considered pure and crystalline. The bodies are exposed to protosolar illumination at a solar nebula 
clearing time, assumed to be $t_{\rm c}=3\,\mathrm{Myr}$ after CAI.

\begin{figure}
\scalebox{0.4}{\includegraphics{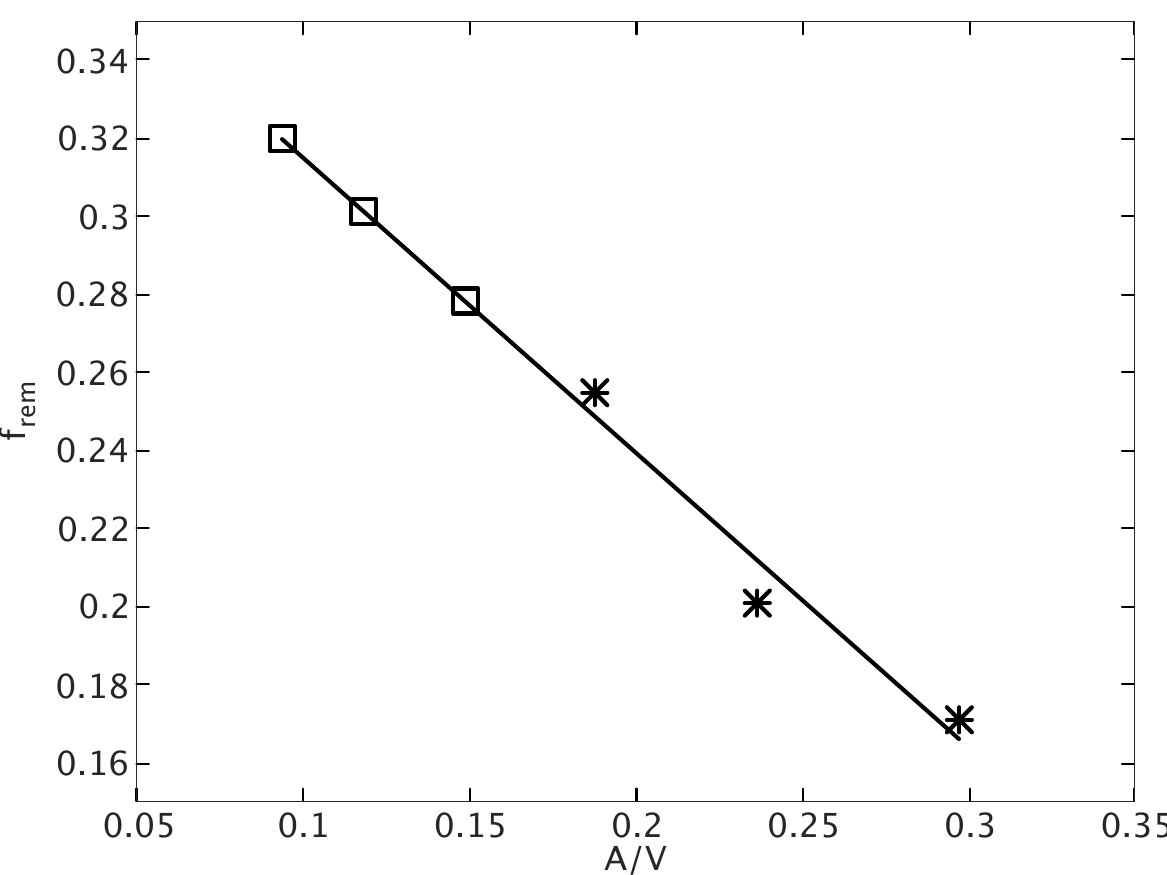}}
     \caption{Zone~\#1 at $15$--$20\,\mathrm{au}$, prior to catastrophic collisions: the remaining fraction $f_{\rm rem}$ of $\mathrm{CO_2:CO}$ mixtures at the 
time of impact $t_{10}$, as function of the surface area to volume ratio $A/V$. The $f_{\rm rem}$ factors for the three largest 
$A/V$ values are properly evaluated through \textsc{nimbus} simulations (asterisks). Those are used to define a linear LSQF 
fit that extrapolates to lower $A/V$ values (bigger body diameters). The squares correspond to the $f_{\rm rem}$ estimates 
for $40.3\leq D\leq 64\,\mathrm{km}$ in Table~\ref{tab_z1b}.}
     \label{fig_frem}
\end{figure}

\begin{table}
\begin{center}
\begin{tabular}{||r|r|r|r||}
\hline
\hline
$n$ & $D_{\rm p}\,\mathrm{[km]}$ & $t_{\rm CO}\,\mathrm{[Myr]}$ & $f_{\rm rem}$\\
\hline
12 & 64.0 &  5.290 & $\sim 0.32$\\
11 & 50.8 &  4.552 & $\sim 0.30$\\
10 & 40.3 &  4.024 & $\sim 0.28$\\
9 & 32.0 &  3.668 & $0.255$\\
8 & 25.4 & 3.427 & 0.201\\
7 & 20.2 & 3.273 & 0.171\\
\hline
\end{tabular}
\caption{Zone~\#1 at $15$--$20\,\mathrm{au}$, prior to catastrophic collisions: the $n^{\rm th}$ generation parent of 67P with diameter $D_{\rm p}$ loses its 2 per cent (molar relative water) of 
condensed CO at $t_{\rm CO}$ (note that simulations are initiated at $t=3\,\mathrm{Myr}$). The fraction of the $\mathrm{CO_2:CO}$ mixture still remaining at the time of collision 
$t_{10}$ is denoted by $f_{\rm rem}$ (if proceeded by `$\sim$' then estimated, otherwise properly calculated).}
\label{tab_z1b}
\end{center}
\end{table}

Table~\ref{tab_z1b} shows that it takes $t_{\rm CO}-3=0.3\,\mathrm{Myr}$ for the smallest ($D=20.2\,\mathrm{km}$) body to lose its condensed CO ice, 
while the $\sim 32$ times more voluminous body ($D=64\,\mathrm{km}$) needs $t_{\rm CO}-3=2.3\,\mathrm{Myr}$ to complete the loss. In all cases, the 
loss time--scales are shorter than the collision time--scales (see Table~\ref{tab_z1a}). Naively, one would expect that the loss time scales linearly with initial abundance for a body of 
a given size. That is because a shell at some given distance from the core is being fed with energy at the same rate regardless of the 
CO abundance in the shell, giving rise to the same loss rate as function of CO front depth. Therefore, if the initial CO abundance was 6 per cent instead of 2 per cent, 
emptying the shell should take three times longer. However, when this hypothesis was tested in an actual simulation, it turned out that a $D=20.2\,\mathrm{km}$ body with 6 per cent CO did not 
require $0.82\,\mathrm{Myr}$ for the loss as expected, but merely $0.55\,\mathrm{Myr}$. The reduction to $\sim 2/3$ of the loss time expected from 
simple linear extrapolation can be understood as follows. Initially, the loss from superficial shells of a given volume indeed takes three times longer. But because of this delay, 
more energy has time to be conducted past the CO front to heat the core. At the points in time when exactly half of the CO remains, the core of the 
body with 2 per cent CO has heated to $35\,\mathrm{K}$ but the same--sized body with 6 per cent CO has had time to heat to $43\,\mathrm{K}$. Because of the 
higher temperature of the latter body, its peak CO gas pressure reached $4.8\,\mathrm{Pa}$, whereas the former body only reached $4.0\,\mathrm{Pa}$. 
The higher gas pressure speeds up the gas diffusion rate and cuts the total loss time short of the extrapolated time. Therefore, a loss time obtained by 
scaling linearly with abundance constitutes an \emph{upper limit}. 

With this in mind, one can ask how much condensed CO a body should have had at the time of solar nebula clearing in order to still have some left 
at the time $t_{10}$ of the catastrophic collision. If parts of the waste energy could be consumed by the sublimation of remaining condensed CO, 
that would help preserving CO deposits trapped within $\mathrm{CO_2}$ or amorphous $\mathrm{H_2O}$. Simple linear extrapolation suggests 
that the $D=20.2\,\mathrm{km}$ body would need $\sim 33$ per cent CO, and the $D=64\,\mathrm{km}$ body would need $\sim 11$ per cent CO. In reality, 
the required abundances are even higher, because these amounts would run out before $t_{10}$, as previously explained. Because the total CO budget likely would 
constitute 13--26 per cent at most (of which a substantial fraction needs to be stored in $\mathrm{CO_2}$ and/or $\mathrm{H_2O}$), 
as discussed in section~\ref{sec_model}, not much CO ice should remain at $t_{10}$. This suggests that cooling by condensed CO at the time of collision may not be efficient.

\begin{figure*}
\centering
\begin{tabular}{cc}
\scalebox{0.4}{\includegraphics{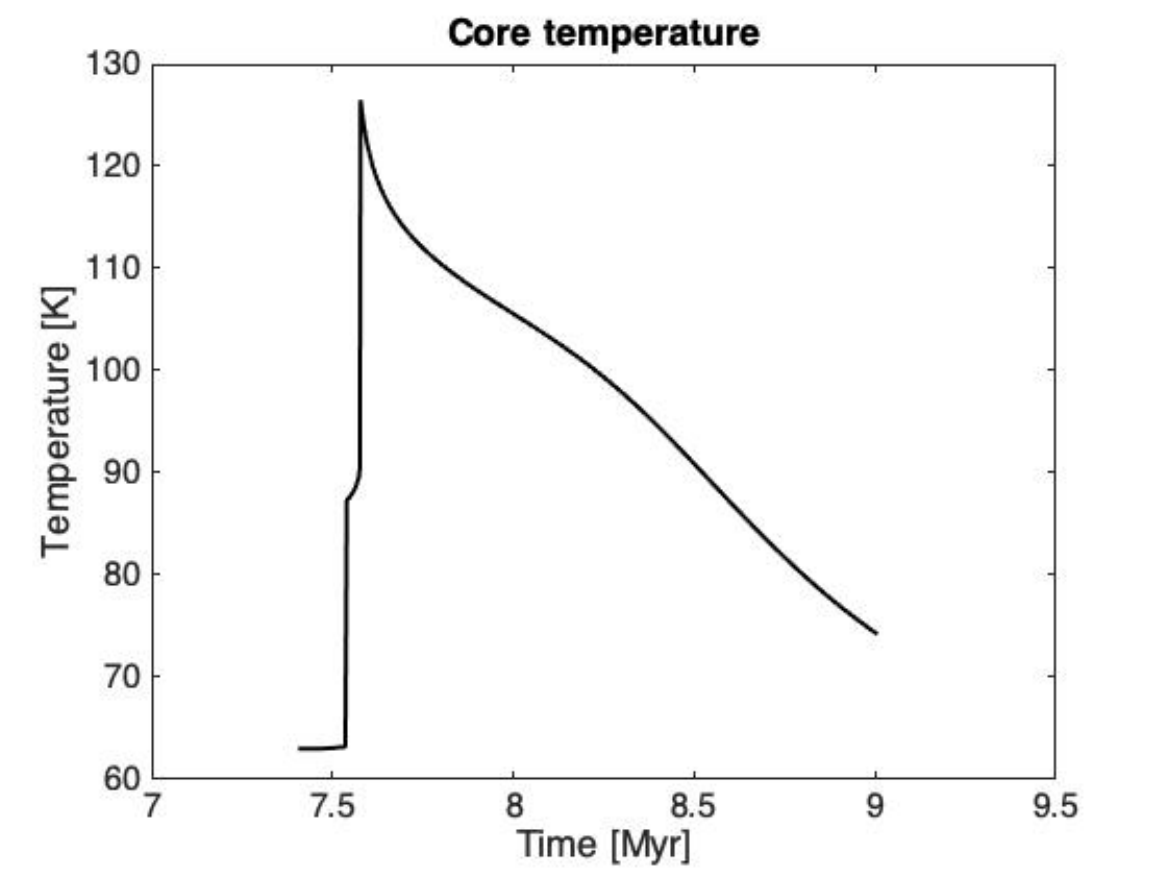}} & \scalebox{0.4}{\includegraphics{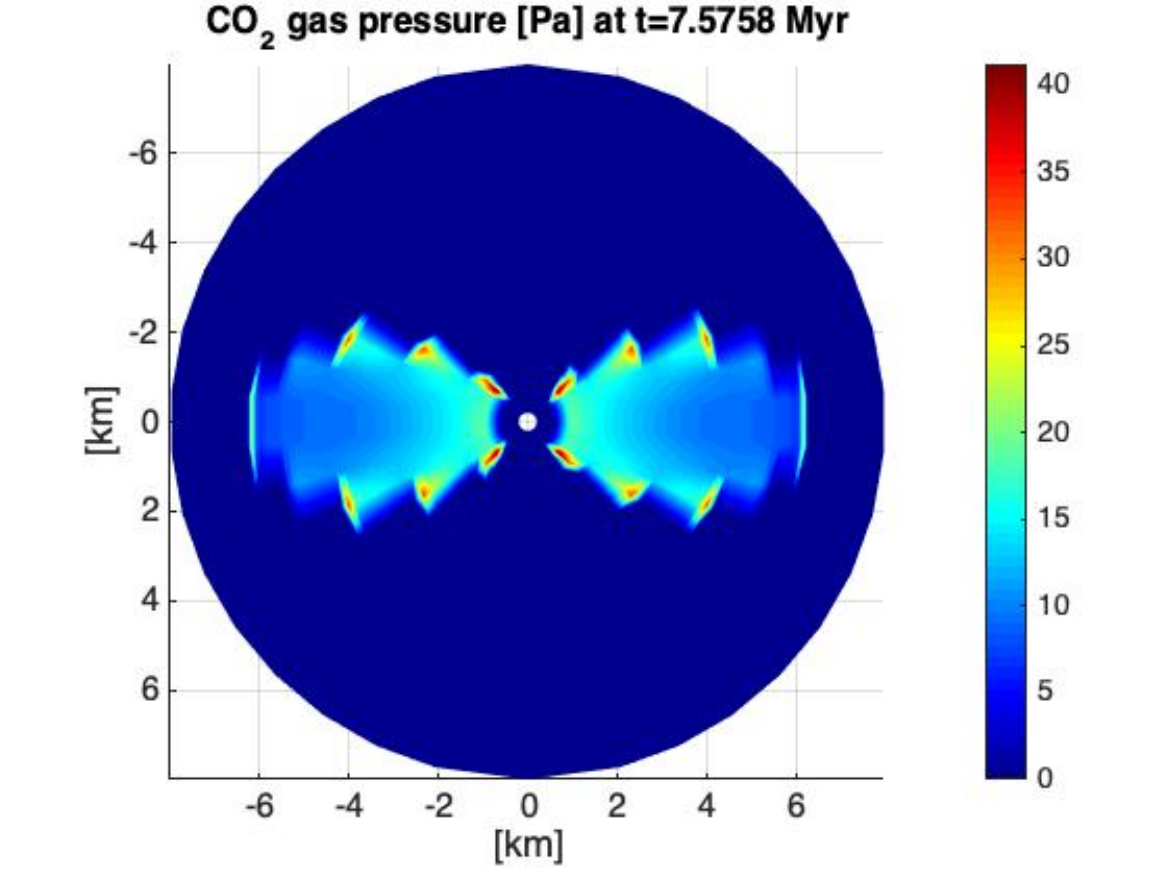}}\\
\end{tabular}
     \caption{A $D=16\,\mathrm{km}$ body at $r_{\rm h}=15\,\mathrm{au}$, with $\mu=4$ refractories/water--ice mass ratio, 5 per cent $\mathrm{CO_2}$ and 2 percent CO (all trapped in 
the water ice that initially is amorphous). Radionuclides are ignored. It formed at $t_{10}=7.5\,\mathrm{Myr}$ as the result of a catastrophic disruption of a $D=20.2\,\mathrm{km}$ parent that deposited 
$Q_{\rm D}^*=6.13\,\mathrm{kJ\,kg^{-1}}$ of waste heat. The collisional heating led to widespread crystallisation, loss of $\sim 96$ per cent of the CO, and to a significant internal displacement of $\mathrm{CO_2}$ 
ice (though virtually no $\mathrm{CO_2}$ was lost to space). \emph{Left:} core temperature as function of time. \emph{Right:} the spatial distribution of the $\mathrm{CO_2}$ vapour pressure near the 
onset of runaway crystallisation. The shape of the peak--pressure region is irregular because of the finite spatial resolution of the model (primarily in the latitudinal dimension).}
     \label{fig_D16a}
\end{figure*}

The simulations show that $\mathrm{CO_2:CO}$ mixtures are not stable at $r_{\rm h}=15\,\mathrm{au}$, at least not with the currently 
assumed activation energy and pre--exponential factor \citep{davidsson21}. Comets formed near $r_{\rm h}=15\,\mathrm{au}$ (either primordially, or 
through catastrophic collisions) would therefore not be able to withstand segregation in the long run. In order for such objects to maintain CO throughout 
their primordial disc lifetime, and to be able to release CO when entering the inner Solar system after deep--freeze storage in the scattered disc beyond 
the Kuiper Belt, or in the Oort Cloud, they need to rely exclusively on the presence of CO--laden amorphous water ice. But importantly, if $\mathrm{CO_2:CO}$ 
mixtures remain at the time of collision, they may help preserving CO--laden amorphous $\mathrm{H_2O}$ ice by absorbing a fraction of the collision energy.

The simulations performed up to $t_{10}$ (completed for the $D=20.2$--$32\,\mathrm{km}$ bodies) show that 17--26 per cent of the 
original amount of $\mathrm{CO_2}$--stored CO is still present at the time of collision. These calculations, which proceed for several million years in 
the simulated world, are very time--consuming and increasingly so with growing diameter. Therefore, simulations were stopped at $t_{\rm CO}$ for 
the  $D=40.3$--$64\,\mathrm{km}$ bodies. Estimates of $f_{\rm rem}$ (i.~e., the fraction of $\mathrm{CO_2:CO}$ mixtures remaining at $t_{10}$) for $D\geq 40.3\,\mathrm{km}$ bodies can be obtained by assuming that 
$f_{\rm rem}\propto A/V$. A smaller body has a relatively large surface area $A$ (capable of absorbing solar radiation and of outgassing CO) compared to its 
volume $V$ (which is proportional to the initial amount of CO), hence less capable of preserving its $\mathrm{CO_2:CO}$ (yielding a lower $f_{\rm rem}$ value), 
than a bigger body. Figure~\ref{fig_frem} illustrates this approximate proportionality at $D=20.2$--$32\,\mathrm{km}$ 
and the fitted line is used to extrapolate $f_{\rm rem}$ values at $D=40.3$--$64\,\mathrm{km}$, given in Table~\ref{tab_z1b}.

\subsubsection{Post--collision evolution at $r_{\rm h}=15\,\mathrm{au}$} \label{sec_results_inner_post}

I start with the least energetic collision that formed a $D_{\rm d}=16\,\mathrm{km}$ daughter from a $D_{\rm p}=20.2\,\mathrm{km}$ parent (details of the collision are in Table~\ref{tab_z1a}). 
A first case (R09\_001C, see Table~\ref{tab_z1c}) had no free CO, clean $\mathrm{CO_2}$, and all CO trapped in amorphous water ice. The purpose of this simulation was to see if the sole 
CO host would survive the collision.

\begin{figure}
\scalebox{0.4}{\includegraphics{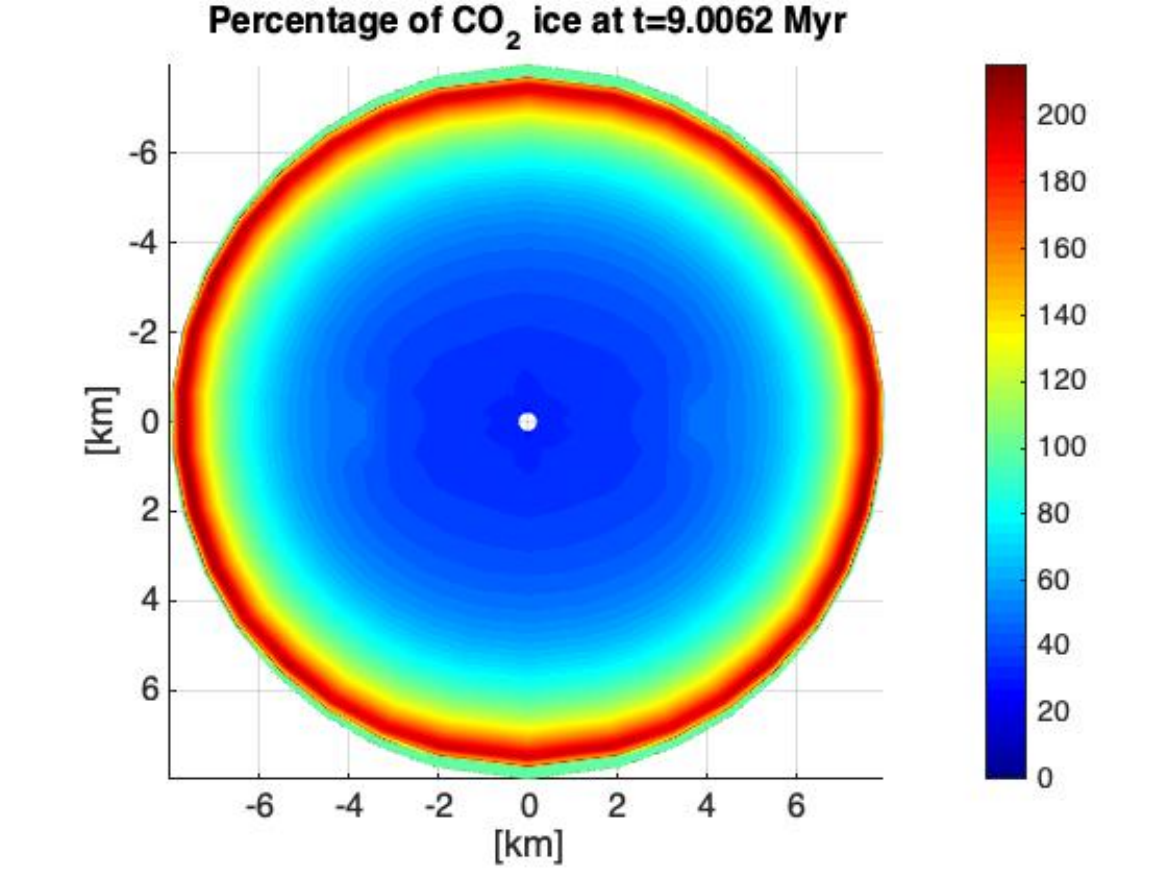}}
     \caption{The final spatial distribution of $\mathrm{CO_2}$ (the initial abundance scaled to 100 per cent) for the same body as shown in Fig.~\ref{fig_D16a}. The $\mathrm{CO_2}$ 
vapour, released through $\mathrm{CO_2}$ ice sublimation triggered by collisional heating and energy release during amorphous water ice crystallisation, has diffused way from 
the core and towards the surface. The low near--surface temperatures led to massive $\mathrm{CO_2}$ vapour recondensation. The $\mathrm{CO_2}$ loss to space is negligible, 
but the internal redistribution of $\mathrm{CO_2}$ ice is substantial.}
     \label{fig_D16b}
\end{figure}

The waste heat released in the collision increased the core temperature from the ambient $63\,\mathrm{K}$ to $87\,\mathrm{K}$, or by $24\,\mathrm{K}$, see Fig.~\ref{fig_D16a} (left panel). 
This initiated a slow partial crystallisation in the equatorial plane close to the surface, where the temperature was highest through a combination of protosolar and collisional heating. During the 
following $43\,\mathrm{kyr}$, the epicentre of partial crystallisation moved in the equatorial plane towards the centre, while releasing heat, which accelerated the crystallisation rate further. 
At that point, runaway crystallisation and strong energy release started. An equatorial torus--like region, located about 1--$6\,\mathrm{km}$ from the core, became completely crystallised. The 
geometry of this region is illustrated in Fig.~\ref{fig_D16a} (right panel) via the spatial distribution of the pressure due to $\mathrm{CO_2}$ vapour, released by vigorous $\mathrm{CO_2}$ ice sublimation. 

In the following $2\,\mathrm{kyr}$, the fully crystallised region spread equatorially to the core and to the surface, and expanded in latitude. The core temperature peaked at $\sim 127\,\mathrm{K}$ 
(see Fig.~\ref{fig_D16a}, left panel, and Table~\ref{tab_z1c}). At most the $\mathrm{CO_2}$ pressure reached $41\,\mathrm{Pa}$. The body then cooled gradually, and the core temperature 
fell below $80\,\mathrm{K}$ about $1.3\,\mathrm{Myr}$ after the collision that formed the body. At that point, merely 4 per cent of the original amount of amorphous ice remained, which 
means that practically all of the CO of the body has been lost to space. 

\begin{table*}
\begin{center}
\begin{tabular}{||l|l|l|l|l|r|l|l|l|l|l||}
\hline
\hline
Model & $n$ & $D_{\rm d}$ & $\mathrm{H_2O:CO}$ & $\mathrm{CO_2:CO}$ & $T_{\rm max}$ & $t_{\rm cool}$ & End tot. a--$\mathrm{H_2O}$ & End core a--$\mathrm{H_2O}$ & End $\mathrm{CO_2:CO}$ & End core $\mathrm{CO_2}$\\
 & & $\mathrm{[km]}$ & [Per cent] & [Per cent] & $\mathrm{[K]}$ & $\mathrm{[Myr]}$ & [Frac] & [Frac]  & [Frac] & [Frac]\\
\hline
R09\_006A & 11 & 50.8 & 2 & 2 & 145.3 & 12.2 & 0.001 & 0.000 & 0.000 & 0.000\\ 
R09\_005A & 10 & 40.3 & 2 & 2& 139.5 & 7.15 & 0.001 & 0.000 & 0.000 & 0.016\\
R09\_004A & 9 & 32.0 & 2 & 2 & 135.2 & 4.56 & 0.002 & 0.000 & 0.000 & 0.064\\
R09\_003A & 8 & 25.4 & 2 & 4 & 128.2 &  2.77 & 0.036 & 0.000 & 0.003 & 0.297\\
R09\_002A & 7 & 20.2 & 2 & 2 & 127.8 & 1.83 & 0.043 & 0.000 & 0.004 & 0.305\\
R09\_002B & 7 & 20.2 & 2 & 3 & 88.2 & 0.28 & 0.990 & 0.881 & 0.007 & 1.000\\
R09\_001C & 6 & 16.0 & 2 & 0 & 126.5 & 1.26 & 0.038 & 0.000 & -- & 0.320\\
R09\_001B & 6 & 16.0 & 2 & 2 & 83.5 & 0.13 & 0.999 & 0.997 & 0.007 & 0.997\\
\hline
\end{tabular}
\caption{Zone~\#1 at $15$--$20\,\mathrm{au}$: properties of daughter nuclei born in the aftermath of a catastrophic collision. All bodies have $\mu=4$, the water 
ice is initially amorphous, the $\mathrm{CO_2}$ molar abundance relative water is 5 per cent, and radiogenic heating is omitted. The ambient steady--state 
core temperature is near $63\,\mathrm{K}$. Columns 1--3 are body identifiers (model tag, generation number $n$, and daughter diameter $D_{\rm d}$). Columns 4--5 are initial conditions (abundance of CO 
trapped in amorphous water ice, and abundance of CO trapped in $\mathrm{CO_2}$, in both cases molar relative water). Columns 6--7 are post--collision core peak temperature, and time 
needed for the core to cool back to $80\,\mathrm{K}$ when collision--related changes have ceased. Columns 8--11 are the surviving fractions of the initial phases and species: amorphous water ice 
(a--$\mathrm{H_2O}$) in total and in the core, unsegregated $\mathrm{CO_2:CO}$ mixtures, and the $\mathrm{CO_2}$ ice remaining at the core (essentially no $\mathrm{CO_2}$ is lost to space).}
\label{tab_z1c}
\end{center}
\end{table*}

The $\mathrm{CO_2}$ vapour released by the crystallisation--induced sublimation gradually diffused towards the surface during the extensive cooling period of the body. 
However, because the near--surface region cooled radiatively very quickly after the collision, it became an impenetrable cold trap for the $\mathrm{CO_2}$ vapour. Large 
amounts of $\mathrm{CO_2}$ re--froze near the surface, so that the body globally lost less than one per mille of its $\mathrm{CO_2}$ ice to space. Figure~\ref{fig_D16b} shows 
the substantial degree of internal $\mathrm{CO_2}$ ice spatial re--distribution. The core abundance is down to 32 per cent of the original amount. In a $\sim 1\,\mathrm{km}$ thick 
near--surface shell, the $\mathrm{CO_2}$ ice abundance has more than doubled with respect to initial values.

It is therefore clear that even a relatively small early--generation ancestor of 67P may have difficulties to preserve its hypervolatiles in a 
collisional cascade, at least at $r_{\rm h}=15\,\mathrm{au}$. This is because the most resilient CO host, amorphous water ice, may crystallise globally. 
However, there is a possibility that some condensed CO may remain at the time of collision (if the original abundance exceeded 33 per cent, 
see Sec.~\ref{sec_results_inner_pre}). More realistically, there may be $\mathrm{CO_2:CO}$ mixtures that did not yet have time to segregate 
(provided that such ices exist and were present). If so, such CO may act as a heat sink and aid the preservation of CO--laden amorphous ice 
by reducing or preventing its crystallisation.

To test this idea, model R09\_001B was considered that is similar to R09\_001C except that the $\mathrm{CO_2}$ ice now contains 2 per cent 
CO (molar, relative water). Note, that this would have required $\sim 12$ per cent trapped CO in $\mathrm{CO_2}$ at the assumed 
solar nebular clearing at $t_{\rm c}=3\,\mathrm{Myr}$, because of the loss reported in Table~\ref{tab_z1b} prior to the collision. Storing CO at a $\mathrm{CO_2:CO}=1:2.4$ 
ratio is unrealistic, but a more satisfactory mixing ratio could be achieved by simply assuming a larger abundance of $\mathrm{CO_2}$ ice. The only important 
property is the absolute abundance of $\mathrm{CO_2}$--trapped CO (in terms of bulk $\mathrm{kg\,m^{-3}}$), because of the latent heat consumed during 
segregation that partially or fully may consume the energy released by crystallisation of amorphous water ice. 

As seen in Table~\ref{tab_z1c}, the core temperature of model R09\_001B peaks at $\sim 84\,\mathrm{K}$. It means that the energy consumption of $\mathrm{CO_2:CO}$ 
segregation has suppressed the net temperature increase of the collisional heating from $24\,\mathrm{K}$ to $21\,\mathrm{K}$. This is sufficient to prevent the 
runaway crystallisation process seen in R09\_001C. Once the body has cooled, $>99$ per cent of the global amorphous water ice (and the CO it contains) is still intact. 
The $\mathrm{CO_2}$ ice has not been mobilised, with almost the entire original abundance remaining at the core. Practically all $\mathrm{CO_2:CO}$ is consumed, so the 
only surviving CO--host of this body is amorphous water ice. This shows that reasonable amounts of $\mathrm{CO_2:CO}$ mixtures may be capable of preserving CO--laden 
amorphous water ice in the $n=7\rightarrow 6$ collision. 

Next, the $n=8\rightarrow 7$ collision (destruction of a $D_{\rm p}=25.4\,\mathrm{km}$ parent and birth of a $D_{\rm d}=20.2\,\mathrm{km}$ daughter) is considered. 
In this case, a 2 per cent CO abundance within $\mathrm{CO_2}$ (model R09\_002A) is not sufficient to prevent crystallisation. This is because the specific waste 
heat is 1.33 times higher than for the  $n=7\rightarrow 6$ collision. Linear scaling of waste heats and CO abundances ($1.33\times 2\approx 2.7$) suggests that 3 per cent 
CO in $\mathrm{CO_2}$ should be sufficient to prevent crystallisation. Model R09\_002B shows that this is not entirely correct. Global crystallisation is indeed prevented, 
but the peak temperature is $88\,\mathrm{K}$ (or $\sim 5\,\mathrm{K}$ above R09\_001B) and the amorphous ice abundance at the very core is down to 88 per cent of 
the initial value. Heat dissipation by conduction is slower for the larger body, which means that the interior of the body is kept above a given temperature for a longer time. 
The degree of crystallisation is not only given by the temperature reached, but also by the duration by which that temperature is maintained. This suggests that it becomes 
progressively more difficult for segregation to prevent crystallisation. 

This is further illustrated by the $n=9\rightarrow 8$ collision (destruction of a $D_{\rm p}=32.0\,\mathrm{km}$ parent and birth of a $D_{\rm d}=25.4\,\mathrm{km}$ daughter). 
With the waste heat increased by another factor 1.33, one could naively expect that $1.33\times 3\approx 4$ per cent CO in $\mathrm{CO_2}$ would be sufficient to prevent 
widespread crystallisation. However, Table~\ref{tab_z1c} shows that this is not the case. Model R09\_003A experiences almost complete crystallisation and CO loss. With 
75 per cent of the $\mathrm{CO_2:CO}$ being segregated at the time of collision (Table~\ref{tab_z1b}), it would imply that the initial CO abundance must exceed $\sim 16$ 
per cent in order to prevent crystallisation. With the largest mixing ratio of $\mathrm{CO_2:CO}=5:1$ seen in laboratory experiments \citep{simonetal19}, it implies a $\mathrm{CO_2}$ abundance 
of at least $\sim 80$ per cent relative water. That is too high compared to the $\mathrm{CO_2/H_2O}=0.32\pm 0.02$ ratio observed in low--mass protostars \citep{pontoppidanetal08}, 
suggesting that there is not enough CO to prevent crystallisation in this type of collisions. In conclusion, a collisional cascade at $15\,\mathrm{au}$ is only allowed for 
$D_{\star}<16\,\mathrm{km}$ bodies if amorphous $\mathrm{H_2O}$ is the only CO host, but could include $D_{\star}\leq 20$--$25.4\,\mathrm{km}$ bodies if sufficiently 
abundant $\mathrm{CO_2:CO}$ is present as well.

The daughters in the range $32\leq D\leq 50.8\,\mathrm{km}$ form sufficiently hot to lose all amorphous water ice and associated CO, except for surface layers that 
are merely a few meters thick. The internal re--distribution of $\mathrm{CO_2}$ is substantial. At depths $\stackrel{<}{_{\sim}} 1.5\,\mathrm{km}$, the $\mathrm{CO_2}$ 
abundance exceeds the initial values (and everywhere below, the abundance has decreased). The $D_{\rm d}=32\,\mathrm{km}$ and $D_{\rm d}=40.3\,\mathrm{km}$ bodies 
still have 6 and 2 per cent of the original $\mathrm{CO_2}$ at their cores, respectively (and the abundances rise gradually towards the initial value at $\sim 1.5\,\mathrm{km}$ depth). 
However, the $D_{\rm d}=50.8\,\mathrm{km}$ body interior has been almost completely deprived of $\mathrm{CO_2}$: less than 100 ppm 
of the original amount of the supervolatile is still present at $> 2\,\mathrm{km}$ depth. The near--surface enhancement factors of $\mathrm{CO_2}$ (going from the smallest to the largest $D=32$--$50.8\,\mathrm{km}$ 
daughter) are 9, 13, and 24 times, respectively. These peaks are located at very shallow depths, ranging from 10--$16\,\mathrm{m}$. I note that this redistribution of $\mathrm{CO_2}$ 
does not lead to any \emph{net} sublimation, and hence no net consumption of energy -- it is therefore not capable of acting as a heat sink, or preventing the amorphous water ice from 
crystallising. However, the $\mathrm{CO_2}$ upward diffusion plays an important role for the transport of heat toward the surface, through its advection. I also note, that if the 
$D=50.8\,\mathrm{km}$  body would disintegrate in a second catastrophic collision, most daughter rubble piles would form from its $\mathrm{CO_2}$--free 
interior. Such daughters, and all smaller bodies down to comet size, formed as the result of a  collisional cascade, would lack both hyper-- and supervolatiles.

\subsection{Mid--disc thermal evolution} \label{sec_results_mid}

\subsubsection{Pre--collision evolution at $r_{\rm h}=23\,\mathrm{au}$} \label{sec_results_mid_pre}

The pre--collision models were run under the same conditions as the $r_{\rm h}=15\,\mathrm{au}$ models in Sec.~\ref{sec_results_inner_pre} 
($\mu=4$ crystalline water ice, 5 per cent $\mathrm{CO_2}$, 2 per cent condensed CO, 2 per cent CO in $\mathrm{CO_2}$, and solar nebular clearing at $t_{\rm c}=3\,\mathrm{Myr}$), 
except that the heliocentric distance was set to $r_{\rm h}=23\,\mathrm{au}$. The loss time scales of condensed CO are reported in Table~\ref{tab_z2b}, and ranges from 
$0.4\,\mathrm{Myr}$ for the $D_{\rm p}=20.2\,\mathrm{km}$ parent, to $3.5\,\mathrm{Myr}$ for the $D_{\rm p}=64.0\,\mathrm{km}$ parent. Compared to the conditions 
at $r_{\rm h}=15\,\mathrm{au}$ (Table~\ref{tab_z1b}), these time scales are 43--51 per cent longer, compared to the solar flux being 57 per cent lower. However, the catastrophic 
collisions are not expected to occur until $23\leq t_{10}\leq 44\,\mathrm{Myr}$, which means that there will be no condensed CO remaining at the time of collisions, for reasonable 
CO abundances. 

Because of the larger heliocentric distance, the temperature does not reach high enough to cause spontaneous $\mathrm{CO_2:CO}$ segregation. Therefore, $\mathrm{CO_2}$ is 
a stable and viable candidate for hypervolatile storage at $r_{\rm} \stackrel{>}{_{\sim}} 23\,\mathrm{au}$.

\begin{table}
\begin{center}
\begin{tabular}{||r|r|r|r||}
\hline
\hline
$n$ & $D_{\rm p}\,\mathrm{[km]}$ & $t_{\rm CO}\,\mathrm{[Myr]}$ & $f_{\rm rem}$\\
\hline
12 & 64.0 &  6.450 & 1.000\\
11 & 50.8 &  5.283 & 1.000\\
10 & 40.3 & 4.486 & 1.000\\
9 & 32.0 &  3.963 & 1.000\\
8 & 25.4 & 3.613 & 1.000\\
7 & 20.2 & 3.391 & 1.000\\
\hline
\end{tabular}
\caption{Zone~\#2 at $20$--$25\,\mathrm{au}$, prior to catastrophic collisions: the $n^{\rm th}$ generation parent of 67P with diameter $D_{\rm p}$ loses its 2 per cent (molar relative water) of 
condensed CO at $t_{\rm CO}$ (note that simulations are initiated at $t=3\,\mathrm{Myr}$). The fraction of the $\mathrm{CO_2:CO}$ mixture still remaining at the time of collision 
$t_{10}$ is denoted by $f_{\rm rem}$.}
\label{tab_z2b}
\end{center}
\end{table}

\subsubsection{Post--collision evolution at $r_{\rm h}=23\,\mathrm{au}$} \label{sec_results_mid_post}

\begin{table*}
\begin{center}
\begin{tabular}{||l|l|l|l|l|r|l|l|l|l|l||}
\hline
\hline
Model & $n$ & $D_{\rm d}$ & $\mathrm{H_2O:CO}$ & $\mathrm{CO_2:CO}$ & $T_{\rm max}$ & $t_{\rm cool}$ & End tot. a--$\mathrm{H_2O}$ & End core a--$\mathrm{H_2O}$ & End $\mathrm{CO_2:CO}$ & End core $\mathrm{CO_2}$\\
 & & $\mathrm{[km]}$ & [Per cent] & [Per cent] & $\mathrm{[K]}$ & $\mathrm{[Myr]}$ & [Frac] & [Frac]  & [Frac] & [Frac]\\
\hline
R06\_010A & 11 & 50.8 & 0 & 2 & 104.1 & 1.77 & -- & -- & 0.005 & 0.994\\ 
R06\_009A & 10 & 40.3 & 0 & 2& 93.5 & 0.89 & -- & -- & 0.009 & 1.000\\
R06\_008A & 9 & 32.0 & 0 & 2 & 88.5 & 0.48 & -- & -- & 0.016 & 1.000\\
R06\_008B$^*$ & 9 & 32.0 & 0 & 2 & 75.0 & -- & -- & -- & 0.117 & 1.000\\
R06\_007A & 8 & 25.4 & 0 & 2 & 80.8 &  0.11 & -- & -- & 0.057 & 1.000\\
R06\_006A & 7 & 20.2 & 0 & 2 & 77.6 & -- & -- & -- & 0.378 & 1.000\\
R06\_006B$^*$ & 7 & 20.2 & 0 & 2 & 68.5 & -- & -- & -- & 0.906 & 1.000\\
R06\_005A & 6 & 16.0 & 0 & 2 & 72.5 & -- & -- & -- & 0.907 & 1.000\\
\hline
\end{tabular}
\caption{Zone~\#2 at $20$--$25\,\mathrm{au}$, post--collision: properties of daughter nuclei born in the aftermath of a catastrophic collision. All bodies have $\mu=4$ (except for models R06\_006B 
and R06\_008B, highlighted with asterisks, that considered $\mu=1$), the water 
ice is crystalline, the $\mathrm{CO_2}$ molar abundance relative water is 5 per cent, and radiogenic heating is omitted. The ambient steady--state 
core temperature is near $52\,\mathrm{K}$. Columns 1--3 are body identifiers (model tag, generation number $n$, and daughter diameter $D_{\rm d}$). Columns 4--5 are initial conditions (abundance of CO 
trapped in amorphous water ice, and abundance of CO trapped in $\mathrm{CO_2}$, in both cases molar relative water). Columns 6--7 are post--collision core peak temperature, and time 
needed for the core to cool back to $80\,\mathrm{K}$ when collision--related changes have ceased. Columns 8--11 are the surviving fractions of the initial phases and species: amorphous water ice 
(a--$\mathrm{H_2O}$) in total and in the core, unsegregated $\mathrm{CO_2:CO}$ mixtures, and the $\mathrm{CO_2}$ ice remaining at the core.}
\label{tab_z2c}
\end{center}
\end{table*}

For the Zone~\#2 post--collision simulations, I assume crystalline $\mathrm{H_2O}$ nominally using $\mu=4$, though two tests with $\mu=1$ were performed, a 5 per cent abundance of $\mathrm{CO_2}$, and a 2 per cent abundance of CO that 
initially is trapped fully within $\mathrm{CO_2}$. The outcome of these simulations are summarised in Table~\ref{tab_z2c}. 

The $n=7\rightarrow 6$ and $n=8\rightarrow 7$ collisions that form $D_{\rm d}=16$--$20.2\,\mathrm{km}$ daughters are not sufficiently energetic to 
segregate all $\mathrm{CO_2:CO}$. After cooling--down, these bodies retain 40--90 per cent of their original content of CO. However, the 
$n=9\rightarrow 8$ and $n=10\rightarrow 9$ collisions that form $D_{\rm d}=25.4$--$32\,\mathrm{km}$ daughters result in severe segregation, 
where merely 2--6 per cent of the $\mathrm{CO_2:CO}$ survives. Based on these simulations, the largest parent body in a mid--disc collisional cascade 
would have $D_{\star}\approx 25\,\mathrm{km}$.

None of the $D\leq 40.3\,\mathrm{km}$ daughters suffer any mobilisation or internal redistribution of $\mathrm{CO_2}$ ice. However, 
the first signs of dislocation of $\mathrm{CO_2}$ from the core to more shallow depths are seen in the $n=12\rightarrow 11$ collision 
that formed the $D=50.8\,\mathrm{km}$ body. 

These results apply when $\mu=4$. However, some comet nuclei may be more ice--rich \citep[e.~g., the dust/water--ice mass ratio may be as low as $\mu\approx 1$ for Comet 67P 
according to][]{davidssonetal22}. Two models (R06\_006B and R06\_008B) were run to test the sensitivity to the $\mu$--value. As mentioned in section~\ref{sec_model} a lowered $\mu$ 
results in: 1) a larger specific heat capacity $c$, thus a smaller temperature change $\Delta T\approx \Delta E/c$ for a given energy change $\Delta E$; 2) a larger bulk density of hypervolatiles for 
a given $\mathrm{CO/H_2O}$ ratio due to a higher concentration of water ice. The first effect is rather modest -- Table~\ref{tab_z2c} shows that $D_{\rm d}=20.2\,\mathrm{km}$ and $32\,\mathrm{km}$ 
daughters become $9\,\mathrm{K}$ and $13\,\mathrm{K}$ cooler, respectively, when $\mu=1$ instead of $\mu=4$. The second effect is more prominent because there is more $\mathrm{CO}$ to 
get rid off (i.~e., more latent energy is required to evacuate all CO). Additionally, the segregation process runs slower at lower temperature. For $D_{\rm d}=20.2\,\mathrm{km}$ the amount of 
surviving $\mathrm{CO_2:CO}$ increases from 40 to 90 per cent. For $D_{\rm d}=32\,\mathrm{km}$ the number goes up from 2 to 12 per cent (perhaps qualifying it as a comet ancestor). 
Combined, these effects would adjust the tentative limiting diameter in a collisional cascade from $D_{\star}\approx 25\,\mathrm{km}$ to $D_{\star}\approx 32\,\mathrm{km}$. 
Thus, a drastic reduction of $\mu$ means that the $D_{\star}$--estimate increases by at most one size class (on the currently considered grid).

\subsection{Outer disc thermal evolution} \label{sec_results_outer}

\subsubsection{Pre--collision evolution at $r_{\rm h}=30\,\mathrm{au}$} \label{sec_results_outer_pre}

The pre--collision models at $r_{\rm h}=30\,\mathrm{au}$ were run under the same conditions as previously (sections~\ref{sec_results_inner_pre} and \ref{sec_results_mid_pre}), i.~e., 
$\mu=4$ crystalline water ice, 5 per cent $\mathrm{CO_2}$, 2 per cent condensed CO, 2 per cent CO in $\mathrm{CO_2}$, and solar nebular clearing at $t_{\rm c}=3\,\mathrm{Myr}$. 
The loss time scales of condensed CO are reported in Table~\ref{tab_z3b}, and grows from $0.6\,\mathrm{Myr}$ for the $D_{\rm p}=20.2\,\mathrm{km}$ parent, 
to $6.0\,\mathrm{Myr}$ for the $D_{\rm p}=64.0\,\mathrm{km}$ parent. Compared to the conditions at $r_{\rm h}=15\,\mathrm{au}$ (Table~\ref{tab_z1b}), these time scales are 2.1--2.6 times longer. With 
catastrophic collisions statistically expected to take place at $39\leq t_{10}\leq 71\,\mathrm{Myr}$ (see Table~\ref{tab_z3a}), there will be no condensed CO remaining at that time. 
There is no illumination--driven $\mathrm{CO_2:CO}$ segregation at these distances (at least not when the clearing time is as late as $t_{\rm c}=3\,\mathrm{Myr}$).

\begin{table}
\begin{center}
\begin{tabular}{||r|r|r|r||}
\hline
\hline
$n$ & $D_{\rm p}\,\mathrm{[km]}$ & $t_{\rm CO}\,\mathrm{[Myr]}$ & $f_{\rm rem}$\\
\hline
12 & 64.0 & 9.025 & 1.000\\
11 & 50.8 &  6.649 & 1.000\\
10 & 40.3 & 5.282 & 1.000\\
9 & 32.0 &  4.452 & 1.000\\
8 & 25.4 & 3.913 & 1.000\\
7 & 20.2 & 3.578 & 1.000\\
\hline
\end{tabular}
\caption{Zone~\#3 at $25$--$30\,\mathrm{au}$, prior to catastrophic collisions: the $n^{\rm th}$ generation parent of 67P with diameter $D_{\rm p}$ loses its 2 per cent (molar relative water) of 
condensed CO at $t_{\rm CO}$ (note that simulations are initiated at $t=3\,\mathrm{Myr}$). The fraction of the $\mathrm{CO_2:CO}$ mixture still remaining at the time of collision 
$t_{10}$ is denoted by $f_{\rm rem}$.}
\label{tab_z3b}
\end{center}
\end{table}

\subsubsection{Post--collision evolution at $r_{\rm h}=30\,\mathrm{au}$} \label{sec_results_outer_post}

For the Zone~\#3 post--collision simulations, I assume amorphous $\mathrm{H_2O}$, a 5 per cent abundance of $\mathrm{CO_2}$, and a 4 per cent abundance of CO, 
divided equally between $\mathrm{CO_2}$ and $\mathrm{H_2O}$ (some models only had 2 per cent CO in $\mathrm{H_2O}$).

Because of the stability of $\mathrm{CO_2:CO}$ mixtures at these heliocentric distances, the entire initial abundance of 
CO--laden $\mathrm{CO_2}$ is available to absorb waste heat, thereby shielding potential CO deposits in the even more resilient amorphous water ice. 
The outcome of these simulations are summarised in Table~\ref{tab_z3c}. 

The capability of $D_{\rm d}=16$--$25.4\,\mathrm{km}$ daughters to retain $\mathrm{CO_2:CO}$ increases with respect to $r_{\rm h}=23\,\mathrm{au}$ 
because of the lower ambient temperature at $r_{\rm h}=30\,\mathrm{au}$, pushing the surviving fraction from $\geq 38$ to $\geq 61$ per cent. 
The $D_{\rm d}=32\,\mathrm{km}$ daughter experienced full segregation. A test showed that potential CO deposits within amorphous $\mathrm{H_2O}$ would be  
fully retained in that type of collision, even in the case of having no $\mathrm{CO_2:CO}$ heat sink. 

For a $D_{\rm d}=40.3\,\mathrm{km}$ daughter, having 2 per cent CO in the $\mathrm{CO_2}$ ice also protected against crystallisation. 
But without such protection, substantial crystallisation was obtained, though 15 per cent amorphous water ice survived in the cooler crust. 
If such a nucleus disrupted, its crystalline and (potentially CO--laden) amorphous ices would likely mix in the resulting daughter nuclei, constituting a small but 
not negligible reservoir of CO. The largest considered daughter, with $D_{\rm d}=50.8\,\mathrm{km}$ experienced $\sim 90$ per cent crystallisation even in 
the presence of a $\mathrm{CO_2:CO}$ heat sink, and would likely lose all amorphous ice and associated CO if no $\mathrm{CO_2:CO}$ was present. It is also noteworthy that 
both $D_{\rm d}=40.3$--$50.8\,\mathrm{km}$ nuclei experienced substantial internal $\mathrm{CO_2}$ ice re--distribution if not protected by 
hosted CO, though the core $\mathrm{CO_2}$ abundance remained just over 40 per cent. 

Based on these simulations, the largest parent body in a collisional cascade in the outer primordial would have $D_{\star}\approx 50$--$64\,\mathrm{km}$, 
if containing CO--laden amorphous water ice, with the exact value depending on whether a buffer exists. However, if the sole carrier of CO is $\mathrm{CO_2}$ 
ice, the largest potential parent body has $D_{\star}\approx 30\,\mathrm{km}$.

\begin{table*}
\begin{center}
\begin{tabular}{||l|l|l|l|l|r|l|l|l|l|l||}
\hline
\hline
Model & $n$ & $D_{\rm d}$ & $\mathrm{H_2O:CO}$ & $\mathrm{CO_2:CO}$ & $T_{\rm max}$ & $t_{\rm cool}$ & End tot. a--$\mathrm{H_2O}$ & End core a--$\mathrm{H_2O}$ & End $\mathrm{CO_2:CO}$ & End core $\mathrm{CO_2}$\\
 & & $\mathrm{[km]}$ & [Per cent] & [Per cent] & $\mathrm{[K]}$ & $\mathrm{[Myr]}$ & [Frac] & [Frac]  & [Frac] & [Frac]\\
\hline
R11\_006A & 11 & 50.8 & 2 & 2 & 126.5 & 7.67 & 0.086 & 0.000 & 0.007 & 0.421\\ 
R11\_005A & 10 & 40.3 & 2 & 2& 79.1 & -- & 1.000 & 1.000 & 0.002 & 1.000\\
R11\_005B & 10 & 40.3 & 2 & 0 & 127.0 & 5.29 & 0.146 & 0.000 & -- & 0.411\\
R11\_005C$^*$ & 10 & 40.3 & 2 & 2 & 79.4 & -- & 0.998 & 0.004 & 1.000 & 1.000\\
R11\_004A & 9 & 32.0 & 2 & 2 & 74.0 & -- & 1.000 & 1.000 & 0.010 & 1.000\\
R11\_004B & 9 & 32.0 & 2 & 0 & 77.9 & -- & 1.000 & 1.000 & -- & 1.000\\
R11\_003A & 8 & 25.4 & 2 & 2 & 70.8 &  -- & 1.000 & 1.000 & 0.607 & 1.000\\
R11\_002A & 7 & 20.2 & 2 & 2 & 65.9 & -- & 1.000 & 1.000 & 0.962 & 1.000\\
R11\_001A & 6 & 16.0 & 2 & 2 & 64.4 & -- & 1.000 & 1.000 & 1.000 & 1.000\\
\hline
\end{tabular}
\caption{Zone~\#3 at $25$--$30\,\mathrm{au}$: properties of daughter nuclei born in the aftermath of a catastrophic collision. All bodies have $\mu=4$, the water 
ice is initially amorphous, the $\mathrm{CO_2}$ molar abundance relative water is 5 per cent, and radiogenic heating is omitted. Model R11\_005C (highlighted with an asterisk) was run 
with a 20 times lower heat conductivity than other models. The ambient steady--state 
core temperature is near $47\,\mathrm{K}$. Columns 1--3 are body identifiers (model tag, generation number $n$, and daughter diameter $D_{\rm d}$). Columns 4--5 are initial conditions (abundance of CO 
trapped in amorphous water ice, and abundance of CO trapped in $\mathrm{CO_2}$, in both cases molar relative water). Columns 6--7 are post--collision core peak temperature, and time 
needed for the core to cool back to $80\,\mathrm{K}$ when collision--related changes have ceased. Columns 8--11 are the surviving fractions of the initial phases and species: amorphous water ice 
(a--$\mathrm{H_2O}$) in total and in the core, unsegregated $\mathrm{CO_2:CO}$ mixtures, and the $\mathrm{CO_2}$ ice remaining at the core.}
\label{tab_z3c}
\end{center}
\end{table*}

The effective heat conductivities applied in this work \citep[obtained by correcting laboratory--measured heat conductivities of compacted materials for porosity, using the method of][]{shoshanyetal02} 
result in a thermal inertia that may be high compared to comet material (see section~\ref{sec_model}). To test the importance of that assumption model R11\_005C considered a 20 times lower 
Hertz factor, which resulted in a thermal inertia ranging $13\leq\Gamma\leq 49\,\mathrm{J\,m^{-2}\,K^{-1}\,s^{-1/2}}$ for the coldest and hottest regions of the model body (for most of the interior 
$\Gamma\approx 30\,\mathrm{J\,m^{-2}\,K^{-1}\,s^{-1/2}}$). This was done for the daughter of a $D_{\rm p}=40.3\,\mathrm{km}$ parent, to see if it could be pushed into crystallisation and complete 
CO loss, despite a 2 per cent $\mathrm{CO_2:CO}$ buffer. Because impact heating is practically instantaneous, the effective heat conductivity has no effect on the temperature reached just after the collision. 
There is, however, a strong effect on the cooling time scale. Model R11\_005C was heated to $79.1\,\mathrm{K}$ by the impact (identical to model R11\_005A). Over the following $5.68\,\mathrm{Myr}$ the 
core temperature increased another $0.3\,\mathrm{K}$ because of (an extremely low level of) crystallisation. At that point the core temperature started to fall, because the outer regions of the body had 
cooled down sufficiently to allow for net energy dissipation from the core. The simulation was stopped $590\,\mathrm{kyr}$ later, when the temperature had dropped by $0.015\,\mathrm{K}$. 
In model R11\_005A (having the nominal heat conductivity) the same level of cooling was completed in $56\,\mathrm{kyr}$. Model R11\_005C spent $\sim 100$ times longer at peak temperature 
than model R11\_005A, yet the only consequence was a 0.2 per cent larger loss of amorphous water ice. For this reason, bodies that are pushed close to the edge of global crystallisation may tip over 
if the effective heat conductivity is sufficiently small. However, that would be relevant only in a rather small $T_{\rm max}$ interval, separating bodies that are too cold to crystallise, and the 
ones for which crystallisation is unavoidable, regardless of heat conductivity.  It does not seem that the choice of heat conductivity has a major influence on the $D_{\star}$ estimate.

\section{Discussion} \label{sec_discussion}

We know that small ($D \stackrel{<}{_{\sim}}10\,\mathrm{km}$) comet nuclei contain substantial amounts of CO 
and measurable quantities of other hypervolatiles. There is also mounting evidence that clean CO ice is evacuated on short ($<1\,\mathrm{Myr}$) 
time--scales from primordial disc objects of that size, prior to relocation to the current distant and cooler comet reservoirs (see section~\ref{sec_intro}). 
However, we do not know if the remaining CO is stored in $\mathrm{CO_2}$ ice, or in the more resilient amorphous $\mathrm{H_2O}$, or 
in both. If a cascade of catastrophic collisions took place in the primordial disc, we know that such a cascade could not have 
involved targets heated to the point of substantial segregation and/or crystallisation, because then the last CO would be lost. Hypervolatiles that may 
survive (within $\mathrm{CO_2}$ and/or amorphous $\mathrm{H_2O}$) near the rapidly cooling surfaces of the first daughter would be mixed 
into the interior during subsequent catastrophic collisions and eventually be lost as well. In this paper, I have attempted to determine the critical target diameter $D_{\star}$ (the starting point 
of admissible collisional cascades) and how it changes with heliocentric distance. I first discuss nominal results, and then comment on their sensitivity to composition and 
physical parameters.

The most restrictive scenario is that amorphous water ice does not exist, and that all hypervolatiles therefore necessarily are 
locked within $\mathrm{CO_2}$ ice until segregation takes place. According to \textsc{nimbus} simulations, segregation is 
spontaneous at $r_{\rm h}=15\,\mathrm{au}$, and $\mathrm{CO_2:CO}$ mixtures only survive near the cold 
poles at $<1$ per cent levels compared to initial abundances. With polar axis orientations likely changing over time, especially if there 
is supposed to be some collisional activity, that last CO would soon be gone as well. However, $\mathrm{CO_2:CO}$ is completely 
stable at $r_{\rm h}=23\,\mathrm{au}$, suggesting that there would be a transition zone somewhere between $15 \stackrel{<}{_{\sim}} r_{\rm h}\stackrel{<}{_{\sim}} 23\,\mathrm{au}$, 
where the CO abundance increases from zero to high values. Unless we accept the idea that some highly active comets are  
CO--free (yet to be identified observationally), it means that the inner edge of the primordial disc should have been located farther 
from the Sun than $15\,\mathrm{au}$, beyond the `segregation line'. If so, such a relatively distant primordial disc might favour a 
late giant planet dynamic instability (see section~\ref{sec_intro}). In that scenario, relatively $\mathrm{CO}$--poor comets \citep[such as 8P/Tuttle, 73P/Schwassmann--Wachmann~3, 103P/Hartley~2, 
C/1999~S4~LINEAR, and C/2000~WM1~LINEAR, with 0.2--0.7 per cent CO relative water;][]{ahearnetal12} may have 
originated close to that inner edge. Farther from the Sun, more CO--rich bodies would form \citep[such as 22P/Kopff, 88P/Howell, and C/2008 Q3 Garradd, with 
$\geq 20$ per cent CO relative water;][]{ahearnetal12}. The survival of such deposits would place constraints on the largest possible parent 
body that could participate in a collisional cascade: $D_{\star}\leq 25\,\mathrm{km}$ at $r_{\rm h}=23\,\mathrm{au}$, growing to $D_{\star}\leq 32\,\mathrm{km}$ at $r_{\rm h}=30\,\mathrm{au}$.

A second possibility is that CO--laden $\mathrm{CO_2}$ does not exist, which means that the CO carrier necessarily has to be amorphous water ice. 
Such ice has long--term stability at $r_{\rm h}=15\,\mathrm{au}$ (when only considering protosolar heating). In the inner regions of the primordial disc, the 
largest acceptable parent in a collisional cascade has $D_{\star}< 16\,\mathrm{km}$. At $r_{\rm h}=30\,\mathrm{au}$ that threshold has increased to $D_{\star}\leq 50\,\mathrm{km}$. 
The crystallisation of amorphous water ice releases additional heat. Combined with collisional heating, it mobilises $\mathrm{CO_2}$ that diffuses from the core closer to 
the surface, where it recondenses. In the current work, $\mathrm{CO_2}$ therefore does not escape to space and therefore does not consume energy through net sublimation, 
though it participates in cooling the bodies through its advection. The current paper ignores the presence of frequent cratering, shape--changing, and sub--catastrophic collisions that 
may cause temporary (or steady--state) heating of the near--surface layers. Such heating may prevent $\mathrm{CO_2}$ recondensation, and allow for loss to space. Such 
collision--driven $\mathrm{CO_2}$ loss will be the topic of a forthcoming paper. 

If CO is carried both by $\mathrm{CO_2}$ and by amorphous water ice, the former acts as a heat sink buffer that may protect the latter. This pushes the 
upper size of parents commonly participating in collisional cascades to $D_{\star}\leq 20$--$25\,\mathrm{km}$ at $r_{\rm h}=15\,\mathrm{au}$, 
and to $D_{\star}\leq 50$--$64\,\mathrm{km}$ at $r_{\rm h}=30\,\mathrm{au}$.

At $r_{\rm h}=15\,\mathrm{au}$ (assuming the primordial disc extended that close to the protosun), it is therefore necessary to limit substantial collisional 
destruction of bodies to the ones with  $D_{\star}\leq 16$--$25\,\mathrm{km}$, depending on composition. For the primordial disc population size used in 
section~\ref{sec_prep_coll_environ}, this means that the lifetime of the inner primordial disc 
cannot have been much more than 7--$9\,\mathrm{Myr}$. At $r_{\rm h}=30\,\mathrm{au}$, the largest parents would have had $D_{\star}\leq 32$--$64\,\mathrm{km}$, 
again depending on composition. That would be possible if the lifetime of the primordial disc did not exceed 50--$70\,\mathrm{Myr}$ at such distances. 
If the lifetime in reality was longer, the population size need to be proportionally smaller. For example, a $\sim 450\,\mathrm{Myr}$ lifetime consistent with 
the Late Heavy Bombardment, would still be possible as long as the population size is reduced by a factor 6--60 relative to current estimates \citep{brassermorbidelli13,morbidellirickman15,rickmanetal15}. 

It should also be pointed out that if the primordial disc lifetime was $\sim 10\,\mathrm{Myr}$ or shorter, at most 10 per cent of the 
$D_{\rm p}=5\,\mathrm{km}$ ($n=1$) nuclei would be collisionally disrupted in the outer half (23--$30\,\mathrm{au}$) of the primordial disc (Tables~\ref{tab_z2a} and \ref{tab_z3a}). 
If that is the case, a majority of the current 67P--type nuclei would be primordially formed bodies, and not collisional fragments.

According to observations, the cumulative size--frequency distribution of TNOs has a turnover from a relatively steep slope at $D\stackrel{>}{_{\sim}} 100\,\mathrm{km}$, to a more 
shallow slope at smaller sizes \citep{bernsteinetal04,fraseretal10,fraseretal14}. Some \citep[e.~g.,][]{fraser09} have interpreted the break as a result of collisional 
processing, so that  $D\stackrel{>}{_{\sim}} 100\,\mathrm{km}$ bodies are mostly primordial, while $D\stackrel{<}{_{\sim}} 100\,\mathrm{km}$ bodies are primarily 
collisional fragments and rubble piles. Others \citep[e.~g.,][]{campobagatinandbenavidez12} have suggested that the kink is primordial and related to the 
original formation mechanism of the planetesimals. For example, gravitational collapse of pebble--swarms formed by streaming instabilities lead to a 
cumulative size--frequency distribution with a turnover near $D\approx 100\,\mathrm{km}$ under some circumstances \citep{lietal19} that seem to be regulated 
by turbulent diffusion \citep{klahrandschreiber21}. The current work shows that even for the most resilient bodies (CO is trapped in amorphous $\mathrm{H_2O}$ and a $\mathrm{CO_2:CO}$ 
heat sink is available), collisional depletion of bodies larger than $D_{\star}=25$--$64\,\mathrm{km}$ (depending on distance, $15\leq r_{\rm h}\leq 30\,\mathrm{au}$) is excluded. If the 
only CO--bearer is $\mathrm{CO_2}$, those limits are shifted to $D_{\star}=25$--$32\,\mathrm{km}$. The thermophysical analysis therefore supports a primordial origin of the observed 
turnover in the TNO size--frequency distribution. 

As mentioned in section~\ref{sec_model}, long--lived radiogenic heating has been ignored, the CO and $\mathrm{CO_2}$ abundances may have been underestimated, while $\mu$ and 
$\kappa_{\rm s}(T,\,\psi)$ may be lower in some objects than considered here. In the 
following, the consequences for the $D_{\star}$ estimates are discussed. The error in $D_{\star}$ introduced by ignoring radiogenic heating can be estimated as follows. The specific energy released in comet material by 
long--lived radionuclides \citep[Table~12 in][]{davidsson21} integrated during $t_{\rm cool}$ increases from $\sim 10^3\,\mathrm{J\,kg^{-1}\,t_{\rm cool}^{-1}}$ at small sizes to  
$\sim 10^4\,\mathrm{J\,kg^{-1}\,t_{\rm cool}^{-1}}$ at large ones. This roughly equals the difference in $Q_{\rm D}^*$ when going from one size class to another (Tables~\ref{tab_z1a}--\ref{tab_z3a}). 
This suggests that $D_{\star}$ would be nudged at most one step downwards (in the list of discrete $D_{\rm p}$--values considered here), if long--term radioactivity was accounted for. 

The CO abundance may be as high as $\nu_5=0.13$--$0.26$ \citep{pontoppidanetal08}. This does not necessarily mean that the assumed 2 per cent abundance of CO trapped in $\mathrm{CO_2}$ 
or $\mathrm{H_2O}$ is an underestimate\footnote{At $r_{\rm h}=15\,\mathrm{au}$ there is a substantial loss of $\mathrm{CO_2:CO}$ due to protosolar--driven segregation between the time 
of formation and $t_{10}$. I emphasise that the usage of 2 per cent CO in $\mathrm{CO_2}$ in the post--collision simulations represents what remains at $t_{10}$, i.~e., the initial 
abundance would have been higher. However, at $23\,\mathrm{au}$ and $30\,\mathrm{au}$, where  $\mathrm{CO_2:CO}$ nominally is stable, the 2 per cent indeed represents an assumed initial condition.}, because we do not know the fraction of all CO that initially was in the form of pure ice. However, laboratory experiments indicate 
a maximum mixing ratio of $\mathrm{CO_2:CO}=5:1$ \citep{simonetal19}, suggesting the $\mathrm{CO_2}$ could support at most $\sim 6$ per cent CO (molar, relative to water). 
Other experiments show a maximum mixing ratio of $\mathrm{H_2O:CO}=100:15$ \citep{barnunetal07}, suggesting the trapping efficiency of CO in $\mathrm{H_2O}$ is nearly 
twice as high as in $\mathrm{CO_2}$. If both substances trapped CO at maximum capacity, the initial abundance of pure CO would have been at most 5  per cent (section~\ref{sec_results} shows that 
such CO would have been lost before the catastrophic disruption). If the abundance of trapped CO indeed is underestimated (by factors $\stackrel{<}{_{\sim}} 3$ in $\mathrm{CO_2}$ 
and $\stackrel{<}{_{\sim}} 7$ in $\mathrm{H_2O}$) it has two consequences: 1) the $\mathrm{CO_2:CO}$ segregation heat sink that delays or prevents CO--release from amorphous 
$\mathrm{H_2O}$ may have been more efficient than currently assumed; 2) the energy release during crystallisation may have been lower than currently assumed, because a larger 
fraction of the crystallisation energy is consumed by CO during its release (absorbing 41 instead of 5 per cent of that energy). Firstly, the majority of the collisional specific energy is 
consumed by raising the temperature from the ambient level to the point of segregation onset. Increasing the amount of CO by a factor $\sim 3$ therefore does not have a drastic 
effect on the capability of the body to absorb collisional energy -- that capability increases by just $\sim 27$ per cent. That is similar to the increase in $Q_{\rm D}^*$ when going 
from one size class to another, which typically is $\sim 33$ per cent. This suggests that $D_{\star}$ would be nudged at most one step upwards, if a higher CO--abundance was 
accounted for. A reduction of $\mu=4$ to $\mu=1$ increases the total mass of CO by a factor 2.5 (for a fixed $\mathrm{CO/H_2O}$ ratio), thus having a similar effect.

Secondly, the reduction of the effective crystallisation energy does not prevent the crystallisation process itself, though it may slow it down. However, because the time 
period near peak temperature is long, crystallisation still has time to complete, and $D_{\star}$ remains nearly the same. The main effect is that the post--crystallisation peak temperature is lower. 
That does not affect the CO loss, but may slightly reduce the level of internal $\mathrm{CO_2}$ displacement. Because $\mathrm{CO_2}$ does not cause a net energy consumption, the 
degree of $\mathrm{CO_2}$ displacement and the initial $\nu_6$--value have no effect on $D_{\star}$. Finally, changes to the heat conductivity does not affect the temperature 
reached right after the impact, and hence does not determine whether or not the body enters the modes of segregation or crystallisation. The nominal simulations used a relatively high heat conductivity 
(short cooling time). Bodies that were disqualified based on their large loss of CO would be even less suitable comet ancestors if cooling times were extended by lowering the nominal theoretical 
thermal inertia towards measured values. The question is then if a lowered thermal inertia would significantly push the $D_{\star}$ estimate to smaller sizes. For the body sizes and heat conductivities 
considered here, $t_{\rm cool}$ is a few times $10\,\mathrm{kyr}$ in cases where crystallisation is not triggered, but could be extended to a few times $1\,\mathrm{Myr}$ if substantially lowering the heat conductivity 
(see section~\ref{sec_results_outer_post}). According to \citet{schmittetal89} a body crystallises in $\sim 50\,\mathrm{kyr}$ if reaching $T_{\rm max}\approx 89\,\mathrm{K}$. If the time available 
at high temperature is extended to $\sim 5\,\mathrm{Myr}$ full crystallisation can be achieved if $T_{\rm max}\approx 85\,\mathrm{K}$. A significant reduction of heat conductivity therefore only extends 
the realm of significant CO loss from bodies heated to $\sim 89\,\mathrm{K}$ in collisions, to somewhat smaller bodies that are heated to $\sim 85\,\mathrm{K}$. In summary, the current $D_{\star}$ 
estimates are expected to be pushed at most one step down (long--term radioactivity) or one step up (ice abundances), on the considered grid of body diameters if other model parameters had been applied. 
For objects that simultaneously have elevated hypervolatile abundances and experience long--lived radiogenic heating, the effects would partially cancel.

The current work has focused on the thermophysical evolution of bodies taken place after collisional disruption and gravitational 
re--accumulation. Simple methods have been used to estimate the amount of energy released during a catastrophic collision. 
It is important to compare those estimates with more rigorous ones from continuum mechanics simulations. In recent years, the capabilities to 
model continuum mechanics numerically have evolved to the point where the disruption of highly porous and icy planetesimals can be studied in 
a sophisticated manner. The parameter space is being explored gradually, and studies have thus far 
not been devoted primarily to the catastrophic disruption of intermediate--sized ($20\leq D\leq 64\,\mathrm{km}$) bodies considered here. 
Notably, investigations have focused on sub--catastrophic \citep{jutzietal17,jutzibenz17} or catastrophic \citep{schwartzetal18} disruption of 
small ($D\stackrel{<}{_{\sim}} 4\,\mathrm{km}$) targets, or alternatively, the cratering, sub--catastrophic, or super--catastrophic impacts onto large 
($40\leq D\leq 400\,\mathrm{km}$) bodies \citep{jutziandmichel20,golabekandjutzi21}. This makes it difficult to directly compare the 
levels of energy release in collision codes, with those applied in the current work.

The published cases most similar to the current ones are probably the hyper--catastrophic disruption (at $2.76 Q^*_{\rm D}$) of a $D=50\,\mathrm{km}$ target 
discussed by \citet{jutziandmichel20}, and the sub--catastrophic disruption (at $0.44 Q^*_{\rm D}$) of a $D=40\,\mathrm{km}$ target 
discussed by \citet{golabekandjutzi21}. They present the results in terms of mass fractions (of the material bound to the daughter, and for the unbound ejecta) that reaches 
a given temperature. \citet{jutziandmichel20} focus on conditions in which water may sublimate (taken as a temperature increase $\Delta T\geq 80\,\mathrm{K}$). They found that 
0.01 per cent of the bound material, and 40 per cent of the unbound ejecta would reach such temperatures. However, the major concern in the current work is 
not water but the loss of the hypervolatile CO, which happens at $\Delta T\approx 10\,\mathrm{K}$ if it primarily is stored within $\mathrm{CO_2}$, or at 
$\Delta T\approx 35\,\mathrm{K}$ if the main host is amorphous water ice. For the sub--catastrophic collision onto a $D=40\,\mathrm{km}$ target, 
\citet{golabekandjutzi21} find $\Delta T\approx 10\,\mathrm{K}$ for 5 per cent of the bound material, and for 50 per cent of the unbound ejecta. 
Note, however, that the impactor would need $\sim 2.3$ times more energy to cause a catastrophic disruption, leading to additional heating. 
This can be compared to model R06\_008A in Table~\ref{tab_z2c}, considering the daughter formed when disrupting a $D_{\rm p}=40.3\,\mathrm{km}$ 
parent. Here, a global $\Delta T=37\,\mathrm{K}$ was obtained. Taken at face value, this heating is more substantial than achieved in the 
simulation by \citet{golabekandjutzi21}. 

First, I caution that the two cases may not be directly comparable. Although specific impact energies are similar, \citet{golabekandjutzi21} consider a 
$d_{\rm proj}=7.2\,\mathrm{km}$ projectile, hitting at velocity $3\,\mathrm{km\,s^{-1}}$ with a $45^{\circ}$ 
impact angle. Model R06\_008A instead considers a $d_{\rm proj}=19.8\,\mathrm{km}$ projectile, hitting head--on at $0.44\,\mathrm{km\,s^{-1}}$. 
The first case has  projectile--to--target mass ratio $m_{\rm d}/M_{\rm p}=0.006$, but model R06\_008A has $m_{\rm d}/M_{\rm p}=0.12$. 
\citet{davisonetal10} studied the level of heating in 50 per cent porosity bodies when the total mass (corresponding to a $12.6\,\mathrm{km}$ diameter body) and 
impact velocity ($5\,\mathrm{km\,s^{-1}}$) were held constant, for different $m_{\rm d}/M_{\rm p}$ values. They were interested in 
the mass fraction reaching the dunite solidus \citep[initiation of rock melting at $1373\,\mathrm{K}$;][]{davison10}. They found insignificant melting at $m_{\rm d}/M_{\rm p}=0.001$ 
but 15 per cent rock melting at $m_{\rm d}/M_{\rm p}=0.1$. This is not only due to differences in specific collision energies: a small and fast projectile causes a less wide--spread 
heating than a large and slow projectile that carries the same kinetic energy. \citet{davisonetal10} explain the reason: heating of the target stops when the release wave catches up with the shock wave. Upon impact, 
shock waves travel both into the target and into the projectile. When the shock wave in the projectile reaches its antipodal collision point, this triggers a release wave 
that goes back through the projectile, enters the target, and catches up with the shock wave of the target. The smaller the projectile, the shorter time is needed for stopping the shock 
wave, and the smaller fraction of the target is compacted and heated. It is therefore expected that a R06\_008A--type collision indeed would lead to substantially 
more heating than the seemingly similar case studied by \citet{golabekandjutzi21}.

Second, the model parameters in most models \citep{jutziandasphaug15,jutzietal17,jutzibenz17,schwartzetal18,jutziandmichel20,golabekandjutzi21} are not necessarily appropriate 
for icy planetesimals (judging from laboratory measurements of analogue materials), and currently seem to bias towards very low levels of heating. These problems are 
discussed further in Appendix~\ref{appendix01}. This is not a criticism of these works, but a recognition of the fact that a relatively 
small part of the parameter space in collision modelling has been explored thus far. It would be unfortunate if the scientific community drew general conclusions 
regarding the level of heating experienced by colliding, porous, and icy planetesimals, before a larger range of options have been considered. That is because of 
the consequences that premature conclusions may have on the interpretation of spacecraft observations of comets (e.~g., by \emph{Rosetta} and eventually 
by \emph{Comet Interceptor}), and the implications for the field of cometary science. Understanding if comets are collisional products or primordial bodies is crucial -- it 
determines whether observed properties inform about original formation or subsequent processing, and has implications on the primordial disc mass and lifetime. 
A critical step is to determine whether heavy collisional processing is compatible with the observed presence of abundant hypervolatiles and supervolatiles in 
comet nuclei. Further studies of low--velocity collisions amongst similarly--sized planetesimals with high porosity, for a wider variety of parameters (see Appendix~\ref{appendix01}) 
are urgently needed.

\section{Conclusions} \label{sec_conclusions}

This paper models the thermophysical evolution of porous icy planetesimals in the $16\leq D\leq 64\,\mathrm{km}$ diameter range 
before and after catastrophic collisional disruption. The focus is on the survival of CO, as representative of all hypervolatiles (e.~g., $\mathrm{N_2}$, $\mathrm{O_2}$, 
$\mathrm{CH_4}$, $\mathrm{C_2H_6}$, and noble gases). They are assumed to be stored within $\mathrm{CO_2}$ and/or amorphous $\mathrm{H_2O}$, which means that 
a potential collisional cascade must avoid segregation and/or crystallisation, otherwise hypervolatiles would not have been abundant in comet nuclei. 
That places constraints on the starting point of the cascade (i.~e., the diameter $D_{\star}$ of the largest body being frequently disrupted). This, in turn, 
places constraints on the lifetime of the primordial disc and/or its population size. The main conclusions based on the nominal models are summarised below.

\begin{enumerate}
\item Bodies with $D\stackrel{>}{_{\sim}} 64\,\mathrm{km}$ cannot avoid crystallisation due to 
long--lived radiogenic heating, even when the dust--to--water ice mass ratio is as low as $\mu=1$. 
Because the most resilient CO--host is lost, such bodies cannot have participated in a collisional cascade, because 
it would produce hypervolatile--free comet nuclei. 
\item At $r_{\rm h}=15\,\mathrm{au}$ it takes 0.3--$2.3\,\mathrm{Myr}$ for $D=20.2$--$64\,\mathrm{km}$ bodies to lose 
2 per cent pure CO ice (relative water, when $\mu=4$). If the CO abundance is increased, the loss time--scale is \emph{shorter} than 
expected from linear extrapolation, because longer time means heating to higher core temperatures, and accelerating vapour evacuation. 
The corresponding loss times at $23\,\mathrm{au}$ and $30\,\mathrm{au}$ are 0.4--$3.5\,\mathrm{Myr}$ and 0.6--$6.0\,\mathrm{Myr}$, respectively. 
\item $\mathrm{CO_2:CO}$ mixtures are not stable at $r_{\rm h}=15\,\mathrm{au}$, at least not for the currently applied activation energy. 
However, 17--32 per cent is expected to remain at $t_{10}$ (the point in time when 10 per cent of the bodies of a given size have been collisionally disrupted), 
helping to protect CO--laden amorphous $\mathrm{H_2O}$. The $\mathrm{CO_2:CO}$ mixtures are stable at $r_{\rm h}\geq 23\,\mathrm{au}$ if the protosun is the only heat source.
\item The most resilient nuclei would store CO in amorphous water ice, and additionally have a heat sink in the form of $\mathrm{CO_2:CO}$ mixtures. 
With such nuclei, a collisional cascade could start at $D_{\star}\leq 20$--$25\,\mathrm{km}$ in the inner ($15\,\mathrm{au}$) part of the primordial 
disc, and at $D_{\star}\leq 50$--$64\,\mathrm{km}$ in its outer ($30\,\mathrm{au}$) part.
\item If there is no $\mathrm{CO_2:CO}$ heat sink, crystallisation of amorphous $\mathrm{H_2O}$ and CO loss can only be avoided in 
collisional cascades starting at $D_{\star}< 16\,\mathrm{km}$ at $15\,\mathrm{au}$ and $D_{\star}\leq 50\,\mathrm{km}$ at $30\,\mathrm{au}$. 
\item If $\mathrm{CO_2}$ is the sole carrier of hypervolatiles, such nuclei would only be stable at $r_{\rm h}\stackrel{>}{_{\sim}} 23\,\mathrm{au}$. 
A potential collisional cascade must have started at $D_{\star}\leq 25\,\mathrm{km}$ at $23\,\mathrm{au}$ and at $D_{\star}\leq 32\,\mathrm{km}$ at $30\,\mathrm{au}$. 
\item Nuclei that crystallise not only lose CO, but the $\mathrm{CO_2}$ is redistributed from the core towards the surface. $D_{\rm d}=32$--$50.8\,\mathrm{km}$ daughters 
formed at $r_{\rm h}=15\,\mathrm{au}$ would have factor 9--24 $\mathrm{CO_2}$ elevations in the top $\sim 1.5\,\mathrm{km}$ (peaking at depths 10--$16\,\mathrm{m}$). 
The partially or fully $\mathrm{CO_2}$--free cores are not suitable comet building--block material. 
\item The $D\approx 100\,\mathrm{km}$ break in size--frequency distribution slopes of TNO populations does not seem consistent with the starting--point of a collisional cascade, 
based on the conclusions i, iv--vi. Because $D_{\star}$ does not change significantly with body composition and conductivity, this conclusion is not biased 
by the considered nominal model assumptions.
\item In order to prevent collisional cascades involving targets larger than the limits defined above, assuming nominal Nice--model population levels, 
the primordial disc lifetime could have been at most $\sim 7$--$9\,\mathrm{Myr}$ at $15\,\mathrm{au}$, and at most $\sim 50$--$70\,\mathrm{Myr}$ at $30\,\mathrm{au}$.  
Alternatively, if the primordial disc lifetime was $450\,\mathrm{Myr}$ (as required if invoking association with the Late Heavy Bombardment), the population levels need to 
be 6-60 times lower than currently assumed. 
\end{enumerate}

\section*{Acknowledgements} 

This research was carried out at the Jet Propulsion Laboratory, California Institute of Technology, under a 
contract with the National Aeronautics and Space Administration. The author acknowledges funding from 
NASA grant 106994 / 811073.02.33.02.90 awarded by the Emerging Worlds program.\\

\noindent
\emph{COPYRIGHT}.  \textcopyright\,2023. California Institute of Technology. Government sponsorship acknowledged.

\section*{Data Availability}

The data underlying this article will be shared on reasonable request to the corresponding author.

\bibliography{MN-22-4331-MJ.R1.bbl}

\appendix

\section{Collisional heating} \label{appendix01}

When large planetesimals collide with one another, they experience a shock wave that drastically increases the density, internal energy (temperature), 
and entropy, while fragmenting and accelerating the material. The shock wave is followed by a release wave that decompresses the material 
isentropically, thereby lowering the density and internal energy to post--collision values \citep[e.~g.,][]{davison10}. These processes 
are described in the context of continuum mechanics, and are governed by the conservation equations for mass, energy, and momentum, the equation of state (EOS; relating pressure, 
density, and internal energy), and the constitutive equation (relating stress and strain, including failure criteria, descriptions of fracture propagation, 
and response to damage). This system of equations is solved by collision codes \citep[e.~g.,][]{benzasphaug94}, while the dispersion and gravitational 
reaccumulation of fragments typically are studied with separate $N$--body simulators \citep[e.~g.,][]{richardsonetal09}. Although such complex numerical 
models are necessary to fully capture the complex behaviour of colliding bodies, insights can be gained by studying the underlying physics. 

The purpose of the following investigation is to show that the level of heating achieved in collision codes is highly sensitive to 
assumed model parameters. Furthermore, the purpose is to better understand how representative recent simulations of porous icy planetesimals 
\citep{jutziandasphaug15,jutzietal17,jutzibenz17,schwartzetal18,jutziandmichel20,golabekandjutzi21} are, because the considered parameter ranges 
are rather restricted. I here apply a Helmholtz free energy on the form
\begin{equation} \label{app01}
\begin{array}{c}
\displaystyle F=\left(aE+\frac{A}{\varrho_0}-\frac{2B}{\varrho_0}\right)\ln\frac{\varrho_{\rm s}}{\varrho_0}+A\left(\frac{1}{\varrho_{\rm s}}-\frac{1}{\varrho_0}\right)-\frac{B}{\varrho_{\rm s}}+\frac{B}{\varrho_0^2}\varrho_{\rm s}\\
\\
\displaystyle +\frac{1}{2}bE\ln\left(\frac{E_0\varrho_{\rm s}^2+\varrho_0^2E}{E_0\varrho_0^2}\right)-E.
\end{array}
\end{equation}
The reason for using this form is twofold. First, the pressure of a solid is related to $F$ by
\begin{equation} \label{app02}
P_{\rm s}=\varrho_{\rm s}^2\frac{\partial F}{\partial\varrho_{\rm s}}
\end{equation}
\citep{thompsonandlauson72}, which means that pressure here takes the form
\begin{equation} \label{app03}
P_{\rm s}=\left(a+\frac{b}{\frac{E}{E_0\eta^2}+1}\right)\varrho_{\rm s} E+A\varphi+B\varphi^2,
\end{equation}
i.~e., equals the widely applied \citet{tillotson62} EOS for a solid and compact substance. Here, $\eta=\varrho_{\rm s}/\varrho_0$ (where $\varrho_0$ is the uncompressed density and $\varrho_{\rm s}$ is 
the density under load), $\varphi=\eta-1$, $E$ is the internal energy, and $\{a,\,b,\,E_0,\,A,\,B\}$ are the standard Tillotson material parameters. Secondly, the specific heat capacity of the solid is 
related to $F$ and $E$ by
\begin{equation} \label{app04}
c_{\rm v}=-T\frac{\partial^2 F}{\partial T^2}=\frac{\partial E}{\partial T}
\end{equation}
\citep{thompsonandlauson72}. In the limits $\varrho_{\rm s}\approx\varrho_0$ and $E\ll E_0$ this means that 
\begin{equation} \label{app05}
\frac{\partial c_{\rm v}}{\partial T}\approx \frac{c_{\rm v}}{T}
\end{equation}
which is solved by 
\begin{equation} \label{app06}
c_{\rm v}=\beta T,
\end{equation}
where $\beta$ is a constant. This is attractive, considering that simple substances like water ice typically have specific heat 
capacities that depend linearly on temperature \citep{klinger81}. Because
\begin{equation} \label{app07}
E=\int_0^Tc_{\rm v}(T')\,dT'=\frac{1}{2}\beta T^2,
\end{equation}
it trivially follows that
\begin{equation} \label{app08}
T=\sqrt{\frac{2E}{\beta}}.
\end{equation}
Finally, because entropy is given by 
\begin{equation} \label{app09}
S=-\frac{\partial F}{\partial T}
\end{equation}
\citep{thompsonandlauson72}, it can be related to internal energy as
\begin{equation} \label{app10}
S=-\left((a+b)\ln\frac{\varrho_{\rm s}}{\varrho_0}-1\right)\sqrt{2\beta E}.
\end{equation}

When calculating the properties of a shocked \emph{porous} medium, I follow the procedure described by \citet{jutzietal08}. 
There are four governing equations, starting with the EOS of the solid constituent grains (equation~\ref{app03}). 
Let the distension $\alpha=\varrho_{\rm s}/\rho_{\rm bulk}$ be the ratio between the density of the solid granular constituents and the bulk density of the porous medium, 
both which may change during compression (and call the unloaded values $\varrho_0$, $\rho_{\rm bulk}^0$, and 
$\alpha_0=\varrho_0/\rho_{\rm bulk}^0$). Note, that when porosity has been removed, $\rho_{\rm bulk}=\varrho_{\rm s}$ and $\alpha=1$. Also note that $\varrho_{\rm s}$ may 
either increase or decrease during compression, depending on whether the expansion caused by heating is less or more important than compression due to the load.

\begin{figure}
\scalebox{0.45}{\includegraphics{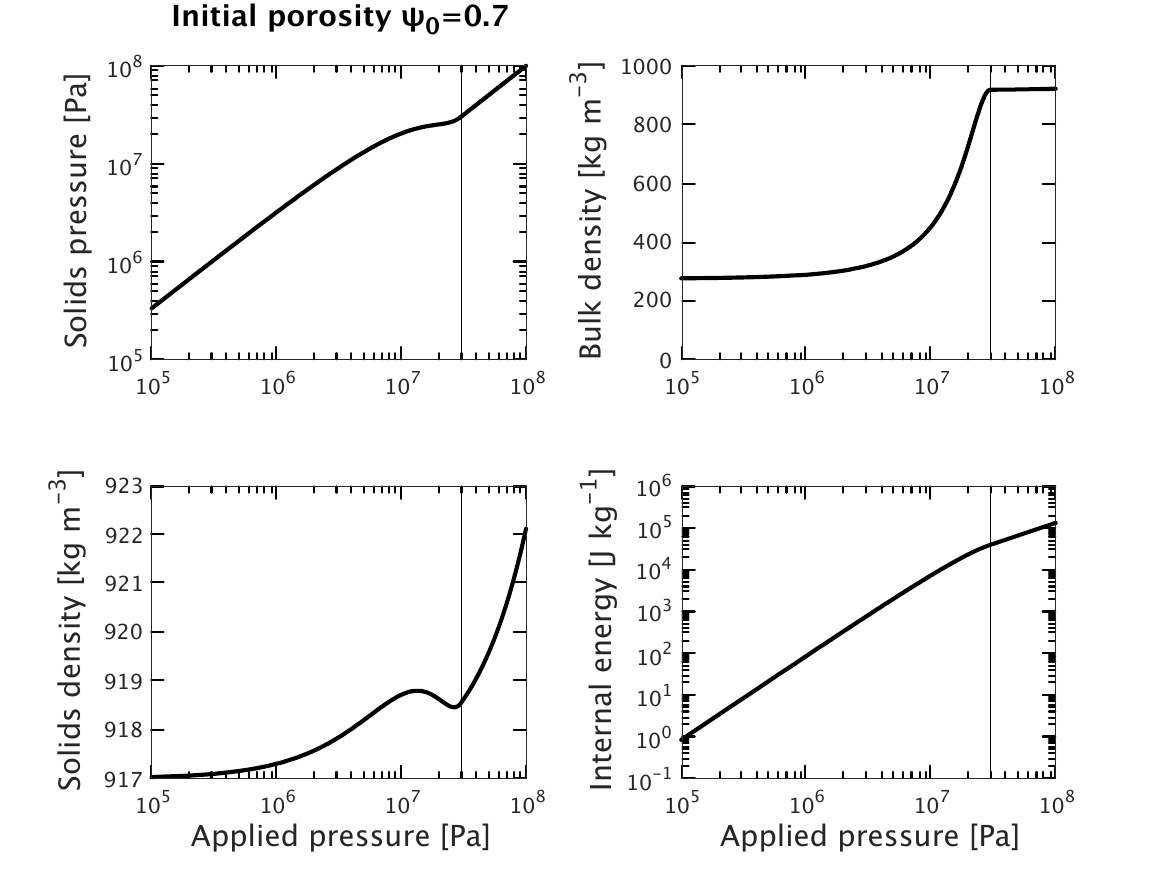}}
     \caption{Behaviour of a porous medium consisting of granular water ice during compression shock (initial porosity $\psi_0=0.7$). In all panels the applied external 
pressure $P$ is shown on the abscissa, and the vertical line marks the applied pore--closing pressure $p_{\rm s}=3\cdot 10^7\,\mathrm{Pa}$ in the $P$--$\alpha$ model. \emph{Upper left:} pressure $P_{\rm s}$ acting on the 
solid grains. \emph{Upper right:} the bulk density $\rho_{\rm bulk}$ of the porous medium. \emph{Lower left:} the density $\varrho_{\rm s}$ of the solid grains. Although the density generally 
increases with $P$, note the local reversal near $P=(1$--$3)\cdot 10^7\,\mathrm{Pa}$ when heating causes expansion that exceeds the level of compression. \emph{Lower right:} the internal energy $E$.}
     \label{fig_compression}
\end{figure}

The relation between the externally applied 
pressure $P$ onto the porous medium, and the pressure $P_{\rm s}$ acting on individual constituent grains are given 
by 
\begin{equation} \label{app11}
P=\frac{1}{\alpha}P_{\rm s}.
\end{equation}
The $P$--$\alpha$ model provides a relation between the external pressure $P$ and the distension,
\begin{equation} \label{app12}
\alpha=1+(\alpha_0-1)\frac{\left(p_{\rm s}-P\right)^2}{\left(p_{\rm s}-p_{\rm e}\right)^2},
\end{equation}
which includes two material parameters: $p_{\rm e}$ being the external pressure at which compression changes from 
elastic to plastic behaviour \citep[I here assume a constant $\alpha=\alpha_0$ in the elastic regime, as done by][]{jutzietal08}, 
and importantly, $p_{\rm s}$ being the external pressure at which \emph{all porosity has been removed} (for simplicity, $p_{\rm s}$ is here 
referred to as the pore--closing pressure). Porosity 
removal has two stages; 1) intact grains slide and rotate with respect to one another until the medium reaches a 
close--packing porosity; 2) grains are deformed and crushed in order to remove porosity entirely. This is an important 
point that I will return to later. Finally, the relation between internal energy, applied pressure, and degree of compression is given by the 
third Hugoniot equation,
\begin{equation} \label{app13}
E=\frac{1}{2}P\left(\frac{1}{\rho_{\rm bulk}^0}-\frac{1}{\rho_{\rm bulk}}\right).
\end{equation}
Equations~\ref{app03}, and \ref{app11}--\ref{app13} have five unknown parameters, and I chose to express $\{P_{\rm s},\,\rho_{\rm bulk},\,\varrho_{\rm s},\,E\}$ as functions 
of the externally applied pressure $P$. Tillotson parameters for water ice from \citet{deniemetal18} are applied (see Table~\ref{tab_tillotson}), and Fig.~\ref{fig_compression} 
illustrates the shocked properties of granular water ice with initial porosity $\psi_0=0.7$ and pore--closing pressure $p_{\rm s}=3\cdot 10^7\,\mathrm{Pa}$ when subjected to 
different levels of external shock pressure $P$ (assuming $p_{\rm e}=0$).

\begin{table}
\begin{center}
\begin{tabular}{||l|r|r||}
\hline
\hline
Parameter & Value & Unit\\
\hline
$a$ & 0.3 & --\\
$b$ & 0.1 & --\\
$E_0$ & $10^7$ & $\mathrm{J\,kg^{-1}}$\\
$A$ & $9.47\cdot 10^9$ & $\mathrm{Pa}$\\
$B$ & $9.47\cdot 10^9$ & $\mathrm{Pa}$\\
$\varrho_0$ & 917 & $\mathrm{kg\,m^{-3}}$\\
\hline
\end{tabular}
\caption{Tillotson parameters for water ice from \citet{deniemetal18}.}
\label{tab_tillotson}
\end{center}
\end{table}

\begin{figure}
\scalebox{0.45}{\includegraphics{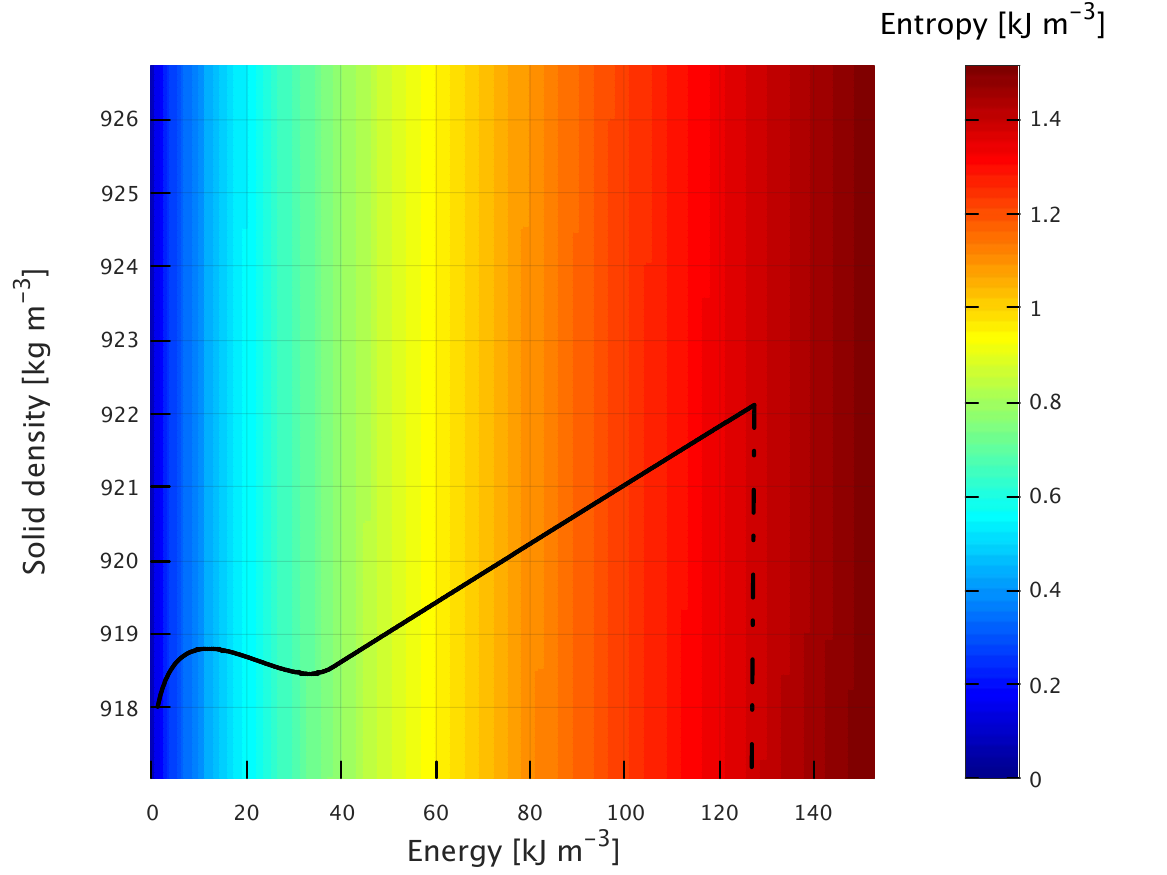}}
     \caption{This figure shows how the entropy (colour coded) changes as increasing external pressure $P$ causes changes in internal energy $E$ and grain density $\varrho_{\rm s}$ 
during the shock wave (solid curve) and during the following release wave (dashed--dotted curve) when $P\rightarrow 0$, $\varrho_{\rm s}\rightarrow\varrho_0$, 
and the entropy remains constant.}
     \label{fig_entropy}
\end{figure}

Equation~\ref{app10} yields an estimate of how the entropy varies as the increasing $P$ causes $\varrho_{\rm s}$ and $E$ to change \citep[assuming $\beta=7.49\,\mathrm{J\,kg^{-1}\,K^{-2}}$;][]{klinger81}. 
Figure~\ref{fig_entropy} shows $\varrho_{\rm s}$ versus $E$ during shock compression as a black solid curve, with the corresponding entropy values $S$ colour--coded in the background. 
When the release wave hits such ice, the medium relaxes back to the unloaded density $\varrho_0$ (I here neglect small deviations because of heat expansion), while the 
entropy remains constant (shown by the dashed--dotted curve). The corresponding reduction in internal energy is merely $\sim 0.4$ per cent. This illustrates the well--known 
principle that the waste heat (net internal energy increase) is large for porous materials, as decompression never is capable of restoring pre--collision porosity \citep{davison10}. 
In the following, I therefore equate the post--relaxation internal energy with the shocked value $E$.

\begin{figure}
\scalebox{0.45}{\includegraphics{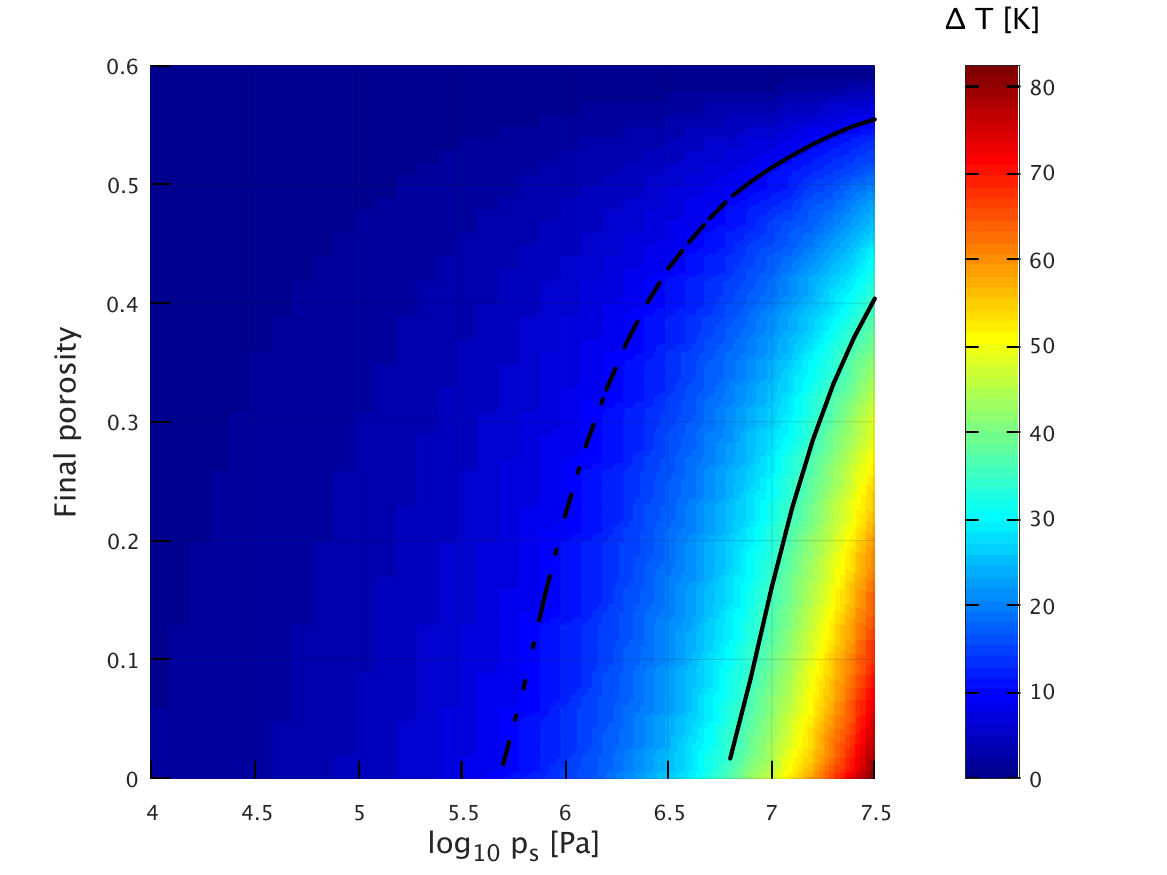}}
     \caption{A porous medium with $\psi_0=0.6$ is considered. Given an assumed pore--closing pressure $p_{\rm s}$ in the $P$--$\alpha$ model, and compression to a given 
post--collision porosity $0\leq\psi\leq\psi_0$, the colour coding shows the temperature increase in the medium. The curves shows the threshold conditions for $\mathrm{CO_2:CO}$ 
segregation (taken as $\Delta T=10\,\mathrm{K}$) and amorphous water ice crystallisation (taken as $\Delta T=35\,\mathrm{K}$), for elevation above the $r_{\rm h}=23\,\mathrm{au}$ ambient temperature of $52\,\mathrm{K}$.}
     \label{fig_dT}
\end{figure}

Applying this method repeatedly for different combinations of $\max(P)$ and $p_{\rm s}$ (assuming different levels of partial compaction, $\max(P)\leq p_{\rm s}$) 
shows how the internal energy (equivalently, temperature, using equation~\ref{app08}) depends on different assumptions about $p_{\rm s}$ for a given 
initial porosity. Figure~\ref{fig_dT} shows the expected $\Delta T$ for different combinations of assumed $p_{\rm s}$ values and post--compression porosities $\psi\leq\psi_0$. 
The figure also shows the post--collision porosities that must be achieved at a given $p_{\rm s}$ to achieve $\mathrm{CO_2:CO}$ segregation (dashed--dotted curve) and 
crystallisation (solid curve) at $r_{\rm h}=23\,\mathrm{au}$.

According to laboratory measurements by \citet{yasuiarakawa09}, porosity removal in granular water ice at $206\,\mathrm{K}$ requires a pressure 
of at least $p_{\rm s}=30\,\mathrm{MPa}$ (or $\log_{10}p_{\rm s}=7.5$), which is near the right edge of Fig.~\ref{fig_dT}. Assuming $\psi_0=0.6$ (close to the macro 
porosities considered in this paper) for such a $P$--$\alpha$ model, it is seen that conditions for segregation are met when compressing to $\psi\approx 0.56$, 
while crystallisation would be initiated when compressing to $\psi\approx 0.4$. Material that experience complete porosity removal is expected to be 
heated by $\Delta T\approx 80\,\mathrm{K}$. The levels of heating seen in the \textsc{nimbus} simulations presented in this paper are similar to 
what one would expect for mild to modest levels of porosity removal, which I take as evidence that the applied $Q_{\rm waste}$ in this paper (admittedly calculated 
with simplified methods) are perfectly reasonable. Note that \citet{durhametal05} report laboratory measurements of ice crush curves that 
\citet{deniemetal18} used to infer $p_{\rm s}=3\cdot 10^8\,\mathrm{Pa}$, which would imply even stronger heating.

The $P$--$\alpha$ models applied in recent work on cometary bodies and other icy planetesimals \citep[e.g.,][]{jutziandasphaug15,jutzietal17,jutzibenz17,schwartzetal18,jutziandmichel20,golabekandjutzi21} 
typically assume $p_{\rm s}$ values in the range $10^4$--$10^6$ (or $4\leq\log_{10}p_{\rm s}\leq 6$), corresponding to the left half of Fig.~\ref{fig_dT}. The temperature 
elevations in this part of parameter space is typically $\Delta T\approx 1\,\mathrm{K}$, even when porosity is removed completely. This is consistent with the statement of 
these authors, that collisions lead to negligible levels of heating. However, it should be recognised that the applied $p_{\rm s}$--values are typical for reaching close--packing in 
porous granular media \citep[e.~g.,][]{guettlereta09}. The work required to achieve such compression is very small, as is the corresponding waste heat. Importantly, such media 
are not compact but typically still have $\sim 40$ per cent porosity. Removing that porosity requires deformation and crushing of grains, which leads to orders--of--magnitude 
higher $p_{\rm s}$--values. That changes the overall behaviour of the material during compression drastically, as seen in Fig.~\ref{fig_dT}: the waste heat becomes substantially 
higher, and temperature elevations measure several times $10\,\mathrm{K}$. 

It is desirable to further study how the assumed values for pore--closing pressure, friction coefficient, cohesion, tensile strength, and sound speed affect the level of heating 
and the partitioning of waste heat between the remaining largest fragment and escaping ejecta. For example, it is known from both laboratory experiments 
\citep[e.~g.,][]{kenkmannetal13,kurosawaandgenda17} and theoretical studies \citep[e.~g.,][]{quintanaetal15,jutzi15} that friction, cohesion, and strength have important 
effects on the level of heating in a collision. In this context is should be remembered that these parameters are largely unknown for real minor icy bodies. Although several 
attempts to measure or infer the strength of cometary material on 67P have yielded values of order 1--$100\,\mathrm{Pa}$ \citep{groussinetal15,groussinetal19,orourkeetal20}, 
there are other measurements that suggest substantially higher (perhaps local) values. \citet{spohnetal15} concluded that the failure of the \emph{Philae}/MUPUS thermal probe 
to penetrate the surface material suggested a compressive strength in excess of $2\,\mathrm{MPa}$. The acoustic waves generated by the MUPUS hammering were sensed by 
SESAME/CASSE, which allowed for an estimate of the Young modulus of $7.2\leq Y\leq 980\,\mathrm{MPa}$ \citep{knapmeyeretal18}. Because the compressive strength of 
ice--dust mixtures typically is 100--400 times lower than $Y$, it could have been anywhere between $18\,\mathrm{kPa}$ and $9.8\,\mathrm{MPa}$ \citep{groussinetal19}. 
If the current population of comet nuclei indeed are the crushed--up and weakened remains of larger planetesimals, we should consider the possibility that those original 
bodies may have been stronger. If so, the start of the collisional cascade should include models with strengths in the $\mathrm{MPa}$ range. The uncertainty in relevant 
parameter values for comet nuclei and other minor icy bodies, combined with the strong dependence of heating on these parameters in 
collision codes, further stresses the importance of widening the exploration of the parameters space in collision modelling.

\bsp	
\label{lastpage}
\end{document}